\documentclass[final,leqno]{siamltex}
\pdfoutput=1
\usepackage{amsmath,amssymb}
\usepackage[sort,compress]{cite}
\usepackage{mathtools}
\usepackage[T1]{fontenc}
\usepackage{graphicx}
\usepackage[section]{algorithm}
\usepackage{algorithmic}
\usepackage{multirow}
\usepackage{makecell}
\usepackage{textcomp}
\usepackage{rotating}
\usepackage{siunitx}
\usepackage{paralist}
\usepackage{color}
\usepackage{url}
\usepackage{colortbl,xcolor}
\usepackage{fancyvrb}
\usepackage{subfigure}

\usepackage[footnotesize,labelfont=bf,textfont=it,up,belowskip=-8pt,aboveskip=5pt]{caption}

\usepackage{animate}

\usepackage[plainpages=false, colorlinks=true,
   citecolor=blue, filecolor=black, linkcolor=blue,
   urlcolor=blue]{hyperref}
\graphicspath{{figures/}}
\definecolor{blue}{rgb}{0,0,1}
\colorlet{lightblue}{blue!50}

\definecolor{myorange}{RGB}{203,96,21}

\newcolumntype{g}{>{\columncolor{gray!30}}r}
\newcolumntype{d}{>{\columncolor{gray!60}}r}
\newcolumntype{p}{>{\columncolor{gray!10}}r}

\usepackage{booktabs}
 
\usepackage{arydshln}

\def\gap{\hspace*{.2in}}

\newcommand{\tabref}[1]{Table~\ref{#1}}
\newcommand{\secref}[1]{\S\ref{#1}}
\newcommand{\algref}[1]{Algorithm~\ref{#1}}

\renewcommand{\b}[1]{{#1}}

\newcommand{\MA}[1]{{\mathcal #1}}
\DeclareMathOperator{\bigO}{\mathcal{O}}
\newcommand{\reals}{\mathbb{R}}

\newcommand{\codenm}[1]{{\tt  #1}}

\newcommand{\msection}[1]{\section{#1}}
\newcommand{\msubsection}[1]{\subsection{#1}}

\newcommand{\ASKIT}{{\sc ASKIT}} 

\newcommand{\bx}{{\b x}} 
\newcommand{\bK}{{\b K}} 
\newcommand{\bKa}{{{\b K}_{\alpha}}} 
\newcommand{\bKas}{{\tilde{{\b K}}_{\alpha}}} 

\newcommand{\SK}{\MA{S}} 
\newcommand{\Sa}{{\SK_\alpha}} 
\newcommand{\ws}{\tilde{w}} 

\newcommand{\XX}{{\MA{X}}} 
\newcommand{\XXa}{\MA{X}_\alpha} 
\newcommand{\XTa}{\MA{T}_\alpha} 

\newcommand{\card}[1]{{|#1|}}

\newcommand{\nl}{{\MA{N}}}            
\newcommand{\nli}{{\MA{N}_i}}        
\newcommand{\nla}{{\MA{N}_\alpha}}   
\newcommand{\rca}{{{\tt r}(\alpha)}} 
\newcommand{\lca}{{{\tt l}(\alpha)}} 
\newcommand{\anc}{\MA{A}} 
\newcommand{\ana}{\anc_\alpha}

\newcommand{\suu}{{\, \cup\, }}          
\newcommand{\sdf}{{\, \setminus\, }}         

\newcommand{\cfun}[1]{{\sc #1}}      

\newcommand{\mcfun}[1]{{\textsc{#1}}} 

\newcommand{\ppl}{{m}} 
\newcommand{\ns}{{s}} 
\newcommand{\ssize}{{\ell}} 

\newcommand{\subK}{{\tilde{K}}}

\newcommand{\dintr}{{d_\mathrm{intr}}}

\title{\ASKIT: Approximate Skeletonization Kernel-Independent
  Treecode in High Dimensions}

\author{William B.~March\footnotemark[1] \and Bo Xiao\footnotemark[1] \and
  George Biros\footnotemark[1]}

\begin{document}
\maketitle
\renewcommand{\thefootnote}{\fnsymbol{footnote}}
\footnotetext[1]{Institute for Computational Engineering and Sciences,
  The University of Texas, Austin, TX}

\begin{abstract}
We present a fast algorithm for kernel summation problems in
high-dimensions. Such problems appear in computational physics,
numerical approximation, non-parametric statistics, and machine
learning. In our context, the sums depend on a kernel function that is
a pair potential defined on a dataset of points in a high-dimensional
Euclidean space. A direct evaluation of the sum scales quadratically
with the number of points. Fast kernel summation methods can reduce
this cost to linear complexity, but the constants involved do not scale well
with the dimensionality of the dataset.

The main algorithmic components of fast kernel summation algorithms
are the separation of the kernel sum between near and far field (which
is the basis for pruning) and the efficient and accurate approximation
of the far field.

We introduce novel methods for pruning and for approximating the far
field.  Our far field approximation requires only kernel evaluations
and does not use analytic expansions.  Pruning is not done using
bounding boxes but rather combinatorially using a sparsified
nearest-neighbor graph of the input distribution.  The time complexity
of our algorithm depends linearly on the ambient dimension. The error
in the algorithm depends on the low-rank approximability of the far
field, which in turn depends on the kernel function and on the
intrinsic dimensionality of the distribution of the points. The error
of the far field approximation does not depend on the ambient
dimension.

We present the new algorithm along with experimental results that
demonstrate its performance. As a highlight, we report results for
Gaussian kernel sums for 100 million points in 64 dimensions, for one
million points in 1000 dimensions, and for problems in which the
Gaussian kernel has a variable bandwidth. To the best of our
knowledge, all of these experiments are prohibitively expensive with
existing fast kernel summation methods.
\end{abstract}


\pagestyle{myheadings}\thispagestyle{plain}\markboth{MARCH, et
  al}{HIGH DIMENSIONAL TREECODE}

\msection{Introduction}\label{s:intro}      

Given a set of $N$ points $\{\bx_j\}_{j=1}^N \in \reals^d$ and weights
$w_j \in \reals$, we wish to compute 
\begin{equation}\label{e:problem}
u_i = u(\bx_i) = \sum_{j=1}^N K(\bx_i, \bx_j) w_j,\quad \forall
i=1\ldots N.
\end{equation}
Here $K()$, a given function, is the \emph{kernel}.\footnote{For
simplicity, we consider only the case in which the input points
$\bx_j$ are both sources and targets.}  Equation~\ref{e:problem} is
the kernel summation problem, also commonly referred to as
an \emph{N-body problem}.  From a linear algebraic viewpoint, kernel
summation is equivalent to approximating ${\b u}={\bK} {\b w}$ where
$\b u$ and $\b w$ are $N$-dimensional vectors and $\bK$ is a $N\times
N$ matrix consisting of the pairwise kernel evaluations. From linear
algebraic viewpoint, fast kernel summations can be viewed as
hierarchical low-rank approximations for $\bK$.

Direct evaluation of the sum requires $\bigO(N^2)$ work. Fast kernel
summations can reduce this cost dramatically. For $d=2$ and $d=3$ and
for specific kernels, these algorithms are extremely efficient and can
evaluate \ref{e:problem} in $\bigO(N\log N)$ work (treecodes) or
$\bigO(N)$ work (fast multipole methods) to arbitrary
accuracy~\cite{cheng-greengard-rokhlin-99,greengard-strain-91}.

The main idea in accelerating \eqref{e:problem} is to exploit
low-rank blocks of the matrix $\bK$. These blocks are related to the
smoothness of the underlying kernel function $K()$, which in turn is
directly related to pairwise similarities between elements of a set,
such as distances between points.  Hierarchical data structures reveal
these low rank blocks by rewriting \eqref{e:problem} as
\begin{equation}\label{e:near-far}
u_i = \sum_{j\in\mathrm{Near}(i)} K_{ij} w_j +
\sum_{j\in\mathrm{Far}(i)}K_{ij} w_j,
\end{equation}
where $\mathrm{Near}(i)$ is the set of points $\bx_j$ whose
contributions cannot be approximated well by a low-rank scheme and
$\mathrm{Far}(i)$ indicates the set of points $\bx_j$ whose
contributions can. The first term is often referred to as
the \emph{near field} for the point $\bx_i$ and the second term is
referred to as the \emph{far field}.  Throughout, we refer to a point
$\bx_i$ for which we compute $u_i$ as a \emph{target} and a point
$\bx_j$ as a \emph{source}.  We fix a group of source points and use
$K$ to represent the interaction%
\footnote{We use the term \emph{interaction} between two
points $\bx_i$ and $\bx_j$ to refer to $K(\bx_i,\bx_j)$.}
 of these source points with distant targets (here we abuse the
notation, since $K$ is just a block of the original matrix). The low
rank approximation used in treecodes and fast multipole methods is
equivalent to a hierarchical low rank factorization of $K$.

The results, algorithms, and theory for the kernels in low dimensions
can be readily extended to any arbitrary dimension, but the constants
in the complexity estimates for both error and time do not scale well
with $d$. For example, the Fast Multipole Method of Greengard and
Rokhlin~\cite{greengard-rokhlin-87} scales exponentially with $d$. We
are interested in developing a fast summation method that can be
applied to points in an arbitrary dimension $d$---provided $K$ has a
low-rank structure.  To clarify, here we are not claiming to be
generically addressing the fast summation for any kernel and any
distribution of points.  As $d$ increases we need to differentiate
between the notion of the \emph{ambient dimension} $d$ and
the \emph{intrinsic dimension} $\dintr$ of the dataset. (For example,
the intrinsic dimension of points sampled on a 3D curve is one.)
Empirically, it has been observed that high-dimensional data are
commonly embedded in some, generally unknown and non-planar, lower
dimensional subspace. An efficient fast summation algorithm must be
able to take advantage of this structure in order to scale well with
both the ambient and intrinsic dimensions. Existing methods do not
exploit such structure.

\textbf{Outline of treecodes.} Roughly speaking, fast summation algorithms for 
\eqref{e:problem} can be
categorized based on 1) how the Near$(i)$ and Far$(i)$ splittings are
defined, and 2) the construction and evaluation of the low rank
approximation of $K_{ij}$ (for $j\in\mathrm{Far}(i)$). First, we
partition the input points using a space partitioning \emph{tree data
structure} (for example $kd$-trees). Then, to evaluate the sum for a
query, we use special traversals of the tree.%
\footnote{Fast Multipole Methods involve more complex logic than a
treecode but they deliver optimal $\bigO(N)$ complexity. Our method
can be extended to behave like an FMM, but we do not discuss the
details in this paper.}
A treecode has the following simple structure (see
Algorithm~\ref{alg_treecode_outline}): for each target point, we
group all the nodes of the tree into {\em Near} and {\em Far}
sets. The interactions from Near nodes are evaluated directly and
those from Far nodes are approximated from their 
approximate representations.  Existing fast algorithms create the
Near/Far groups based on some measure of the distance of a node from
the target point. When a node is far apart or \emph{well-separated}
from a target, the algorithm terminates the tree traversal and evaluates the
approximate interactions from the node to the target point. We refer to this
termination as \emph{pruning} and the distance-based schemes to group
the nodes to Near and Far field as \emph{distance pruning}.

\textbf{The far field approximation.} The far-field approximate
representation for every node has been constructed during a
preprocessing step.  In a companion paper~\cite{march-biros14}, we
review the main methods for constructing far-field approximations, and
we introduce the approximation we use here. We also review this approach in
\secref{s:algorithms}.

\textbf{Shortcomings of existing methods.} In high dimensions 
(\emph{e.g.}~$d>100$), most existing methods for constructing far field
approximations fail because they become too
expensive, they do not adapt to $\dintr$, and distance pruning
fails in high dimensions.  Also, most schemes use analytic
arguments to design the far-field and depend on the type or class of
the kernel.  Although there has been extensive work on these methods
for classical kernels like the Gaussian, other kernels are also used
such as kernels with variable bandwidth that are not shift
invariant~\cite{silverman86}. This observation further motivates the
use of entirely algebraic acceleration techniques
for \eqref{e:problem}.

\subsection*{Contributions} We present "\ASKIT" (Approximate
Skeletonization Kernel Independent Treecode), a fast kernel summation
treecode with a \emph{new pruning scheme} and a \emph{new far field low-rank
representation}.  Our scheme depends on both the decay properties of
the kernel matrix and the manifold structure of the input points.  In
a nutshell our contributions can be summarized as follows:
\begin{itemize}

\item {\em Pruning or near field-far field node grouping.} In \ASKIT{},
pruning is {\em not} done using the usual distance/bounding box
calculations. Instead, we use a combinatorial criterion based on
nearest neighbors which we term \emph{neighbor
pruning}. Experimentally, this scheme improves pruning in high
dimensions and opens the way to more generic similarity
functions. Also, based on this decomposition, we can derive complexity
bounds for the overall algorithm.

\item {\em Far field approximation.} 
Our low rank far field scheme uses an approximate interpolative
decomposition (ID) (for the exact ID
see~\cite{martinsson-rokhlin-tygert07,halko-martinsson-tropp11}) which
is constructed using {\em nearest-neighbor} sampling augmented with
randomized uniform sampling.  Our method enjoys several advantages
over existing methods: it only requires kernel evaluations, rather
than any prior knowledge of the kernel such as in analytic
expansion-based schemes; it can evaluate kernels which depend on local
structure, such as kernels with \emph{variable bandwidths}; and its
effectiveness depends only on the linear algebraic rank of sub-blocks
of the kernel matrix and provides near-optimal (compared to SVD)
compression without explicit dependence on the ambient dimension of
the data. The basic notions for the far field were introduced
in \cite{march-biros14}. Here we introduce the hierarchical scheme,
and evaluate its performance.

\item {\em Experimental evaluation.} One commonly used kernel in statistical
learning is the Gaussian kernel
$K(\bx_i-\bx_j)=\exp(-\|\bx_i-\bx_j\|^2_2/\sigma_j^2)$. We focus our
experiments on this kernel and we test it on synthetic and scientific
datasets. We also allow for a bandwidth that depends on the source or
target point---the variable bandwidth case.  We demonstrate the linear
dependence on the ambient dimension by conducting an experiment with
$d=1000$ (and $\dintr=4$) in which the far field cannot be truncated
and for which we achieve six digits of accuracy and a 20$\times$
speedup over direct $N^2$ evaluation.  On a 5M-point, 18D UCI Machine
Learning Repository \cite{bache-lichman13} dataset, we obtain
25$\times$ speedup. On a 5M-point, 128D dataset, we obtain
2000$\times$ speedup using 4,096 \codenm{x86} cores. Our largest run
involved 100M points in 128D on 8,192 \codenm{x86} cores.

\end{itemize}

In our experiments, we use Euclidean distances and classical binary
space partitioning trees. We require nearest-neighbor information for
every point in the input dataset.  The nearest neighbors are computed
using random projection trees~\cite{dasgupta-freund08} with greedy
search (Section \ref{s:algorithms}). Our implementation combines the Message
Passing Interface (MPI) protocol for high-performance distributed
memory parallelism and the OpenMP protocol for shared memory
parallelism. 

\subsection*{Limitations} The main limitation of our method is that
the approximation rank (the \emph{skeleton size}
in \secref{s:algorithms}) is selected manually and is fixed.  In a
black-box implementation for use by non-experts, this parameter needs
to be selected automatically. The current algorithm also has the
following parameters: the number of approximate nearest neighbor, the
desired accuracy of approximation, the number of sampling points, and
the number of points per box. The nearest neighbors have a more subtle
effect on the overall scheme that can be circumvented with an adaptive
choice of approximation rank. In our experiments, their number is
fixed, but this could also be adaptive. The other parameters are
easier to select, and they tend to affect the performance more than
the accuracy of the algorithm.  Furthermore, the error bounds that we
present are derived for the case of uniform sampling and the analysis
is not informative on how to use the various parameters. More accurate
analysis can be done in a kernel-specific fashion.  Another
shortcoming of the method regards performance optimization. This is a
first implementation of our scheme and it is not optimized. A
fast-multipole variant of this method would also result in $\bigO(N)$
complexity, but we defer this to future work.

Finally, let us mention a fundamental limitation of our scheme. Our
far-field approximation requires that blocks of $K$ have low rank
structure. The most accurate way to compute this approximation is
using the singular value decomposition. There are kernels and point
distributions for which point distance-based blocking of $K$ does not
result in low-rank blocks. In that case, \ASKIT{} will either produce
large errors or it will be slower than a direct sum.  Examples of such
difficult to compress kernels are the (2$d$ or 3$d$) high frequency
Helmholtz kernel~\cite{engquist-ying07} and in high intrinsic
dimensions the Gaussian kernel (for certain bandwidths and point
distributions)
\cite{march-biros14}.

\subsection*{Related work} 
Originally, fast kernel summation methods were developed for problems
in computational physics. The underlying kernels are related to
fundamental solutions of partial differential equations (Green's
functions).  Examples include the 3D Laplace potential (reciprocal
distance kernel) and the heat potential (Gaussian kernel).  Beyond
computational physics, kernel summation can be used for radial basis
function approximation methods. They also find application to
non-parametric statistics and machine learning tasks such as density
estimation, regression, and classification.  Linear inference methods
such as support vector machines \cite{suykens1999least} and dimension
reduction methods such as principal components
analysis \cite{mika1998kernel} can be generalized to kernel
methods \cite{bishop2006pattern}, which in turn require fast kernel
summation.

Seminal work for kernel summations in $d\leq 3$
includes~\cite{greengard-rokhlin-87,cheng-greengard-rokhlin-99}
and~\cite{greengard-strain-91}. In higher dimensions related work
includes \cite{griebel-wissel13,gray-moore01,lee-gray-moore06,morariu-duraiswami08,yang-duraiswami-gumerov03}. One
of the fastest schemes in high dimensions is the improved fast Gauss
transform~\cite{yang-duraiswami-gumerov03,morariu-duraiswami08}. In
all these methods, the low rank approximation of the far field is
based on analytic and kernel-specific expansions. The cost of
constructing and evaluating these expansions scales 
either as $\bigO(c^d)$ (resulting in
SVD-quality errors) or as $\bigO(d^c)$, where $c>0$ is related to the
accuracy of the expansion. Except for very inaccurate
approximations, $c>2$ and thus all of these schemes become extremely
expensive with increasing $d$.  In addition to the expensive scaling
with ambient dimension, the approximations in these methods must be
derived and implemented individually for each new kernel function.

An alternative class of methods is based on a hybrid of analytic
arguments and algebraic approximations.  Examples
include~\cite{ying-biros-zorin-03,fong-darve09,gimbutas-rokhlin-02}. However,
these methods also scale as $\bigO(c^d)$ or
worse~\cite{march-biros14}. 
We also mention methods which rely only on kernel evaluations and use 
spatial decomposition to scale with $\dintr$ in which the far-field 
approximation is computed
on the fly with exact error guarantees \cite{gray-moore01} or approximately 
using Monte Carlo methods \cite{lee-gray08}. However, these methods rely on 
distance-based pruning, which can fail in extremely high dimensions, and the
randomized method is extremely slow to converge. 
Another scheme that only requires kernel
evaluations (in the frequency domain) is \cite{rahimi-recht07}. But
its performance depends on the ambient dimension and on the kernel
being diagonalizable in Fourier space.  For a more extensive
discussion on the far field approximation that we use here and a more
detailed review of the related literature, we refer the reader to our
work~\cite{march-biros14}.

Algebraic methods work with the matrix entries directly. One set of methods
similar to the one presented here.
uses a low-rank matrix decomposition of subblocks of the kernel matrix
\cite{martinsson-rokhlin07}. However, this method still requires interpolation points over a 
bounding surface, which scale poorly with $d$. Another collection of methods,
generally called ``adaptive cross approximation''  methods, also make use
of low-rank factorization of subblocks \cite{bebendorf2000approximation, 
kurz2002adaptive, zhao2005adaptive}. The linear scaling of these 
methods requires the ability to decompose the input points into large, well-
separated sets, which will generally only be possible in low
dimensions.

To the best of our knowledge, all existing treecodes use
distance-based pruning, sometimes augmented with kernel evaluations to
better control the error. Our scheme is the first to introduce an
alternative pruning approach not based on kernel evaluations or
bounding box-based distance calculations.

Since we present experimental results performed in parallel, we also
mention another existing body of work on parallel
treecodes \cite{lee-vuduc-gray13}.  However, our parallel algorithms
are quite different (and our efficiency relies on neighbor
pruning). The specifics will be reported elsewhere since the
parallelization of \ASKIT{} is not our main point here.

We use randomized tree searches to compute approximate nearest
neighbors. There is a significant body of literature on such methods,
but we do not discuss them further since our scheme does not depend on
the details of the neighbor search (although it does depend on the
approximation error, if the nearest neighbors have been computed
approximately.)  Representative works include~\cite{indyk-andoni08}
and \cite{dasgupta-freund08}. We describe the method used in our
experiments in \secref{askit_subsec}.

\msection{Algorithms}\label{s:algorithms}   
\begin{table}[ht!]
\centering
\begin{tabular}{|ll|}
\hline
$N$      & number of points  \\
$K$      & the kernel function or matrix\\
$\kappa$ & number of nearest neighbors per point \\
$\ppl$    & number of points per leaf\\
$\ns$    & number of skeleton points \\
$\XX$    & integer set $\{1,\ldots,N\}$ \\
$\XXa$   & ids of points in node $\alpha$ \\
\hline
$\alpha$ & tree node or simply node \\
$\alpha_i$ & leaf node that contains point $i$\\
$\nli$   & neighbor list for point $i$ \\
$\nla$   & neighbor list for node  $\alpha$ \\
$\ana$   & ancestor list of node $\alpha$ \\
$\rca$   & right child of node $\alpha$ \\
$\lca$   & left child of node  $\alpha$ \\
\hline
\end{tabular}
\caption{Here we summarize main \textbf{notation} used in the text. In
  addition to the information above, we use $\suu$ to indicate the set
  union of two index sets, $\sdf$ the difference of two index sets, and
  $\card{\cdot}$ the number of elements in set. We define
  \cfun{IsLeaf}($\alpha$) to be {\tt true} if $\alpha$ is a leaf.
We also use $w(\MA{I})$ to indicate the components of
vector $w$ determined by an index set $\MA{I}$ and we use a similar
 notation for matrices.}
\label{t:notation}
\end{table}

We now turn to the description of the new algorithm.  We begin by
summarizing the basic structure of a treecode. In a treecode, we first
construct a tree that is used for space-partitioning of the input
points.  We use this term to broadly cover any hierarchical
partitioning of the data set such that nearby (or similar) points are
grouped together. For our purposes, a tree consists of internal nodes
with two or more children and leaf nodes with no children.

Given such a tree, a treecode performs a two-stage computation to
approximate the kernel summations. In the first stage, a bottom-up
tree traversal takes place (also known as the upward pass) in which at
each node we create a low-rank approximation of the far field
generated by all the source points in it. We form these
representations at the leaf nodes, then pass them up to parents and
combine them in a recursive manner. In the second stage, a concurrent
top-down traversal (also known as the downward pass) takes place in
which we use these representations to compute approximate
potentials. That is, for each target point $x$, we traverse the tree
from the top down.  At a node $\alpha$, we apply a {\em pruning
  criterion} to determine whether or not we can approximate the
far-field generated from sources in $a$ and evaluated at $x$.  If we
can approximate it, then we use the low-rank approximation to evaluate
the far field at $x$ and then we \emph{prune} the tree traversal at the node
$a$.  If we cannot approximate it, we recurse and visit the children
of $a$.  If we still cannot prune at a leaf, we evaluate the
contribution of the leaf's points directly (no approximation takes
place).  If by $\subK$ we denote the approximate kernel (meaning that
we use a low rank approximation), a template for a generic treecode is
given by Algorithm~\ref{alg_treecode_outline}. 
 \begin{algorithm}
 \caption{\textbf{Treecode}(Target point $x$, Source tree node $a$)}
 \begin{algorithmic}
 \IF[\emph{using pruning criterion}]{we can approximate $u(x) = K(x,\XXa)w(\XXa)$}  
   \RETURN $u(x) = \subK_a(x)$
 \ELSIF{$a$ is a leaf}
   \RETURN $\sum_{x_j \in a} K(x, x_j) w_j$
 \ELSE
   \RETURN $u(x) = \sum_{a^\prime} \textbf{Treecode}(x, a^\prime)$ for
   all children $a^\prime$ of $a$ 
 \ENDIF
 \end{algorithmic}
 \label{alg_treecode_outline}
 \end{algorithm}
As described, the algorithm results in $\bigO(N \log N)$
complexity. The Fast Multipole Method \cite{greengard-rokhlin-87}
extends this idea by also constructing an \emph{incoming
  representation} which approximates the potentials due to a group of
distant sources at a target point; it results in $\bigO(N)$
complexity.  As described the algorithms has two main technical components: how
do we decide to prune and how do we construct $\subK$?

The vast majority of existing codes use distance-based pruning -- \emph{i.e.}~
they use the minimum distance between $x$ and the bounding box of $a$
relative to the size of the bounding box. If this distance is greater
than a threshold, pruning takes place. This distance can be directly
related to error estimates using the smoothness properties of the
kernel.

\begin{figure}[tp]
        \centering
\subfigure[\textbf{Analytic.}\label{fig_series}]{\includegraphics[width=0.3\textwidth]{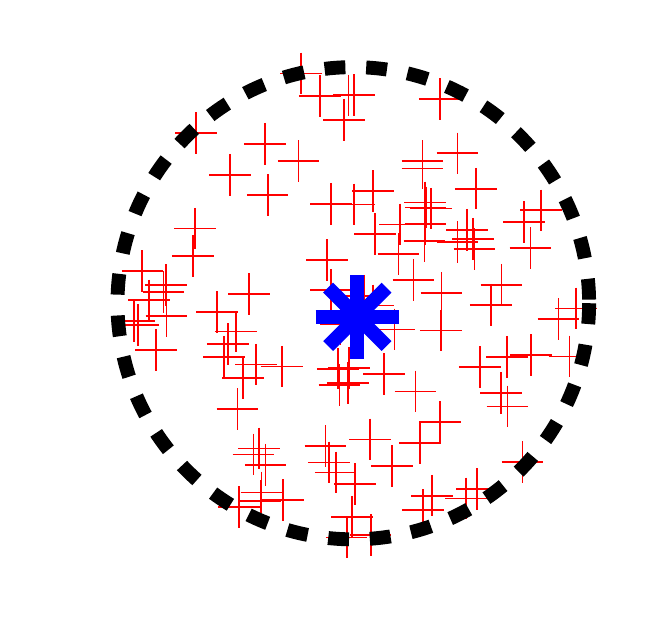}}
        \subfigure[\textbf{Semi-Analytic.}\label{fig_interpolation}]{\includegraphics[width=0.3\textwidth]{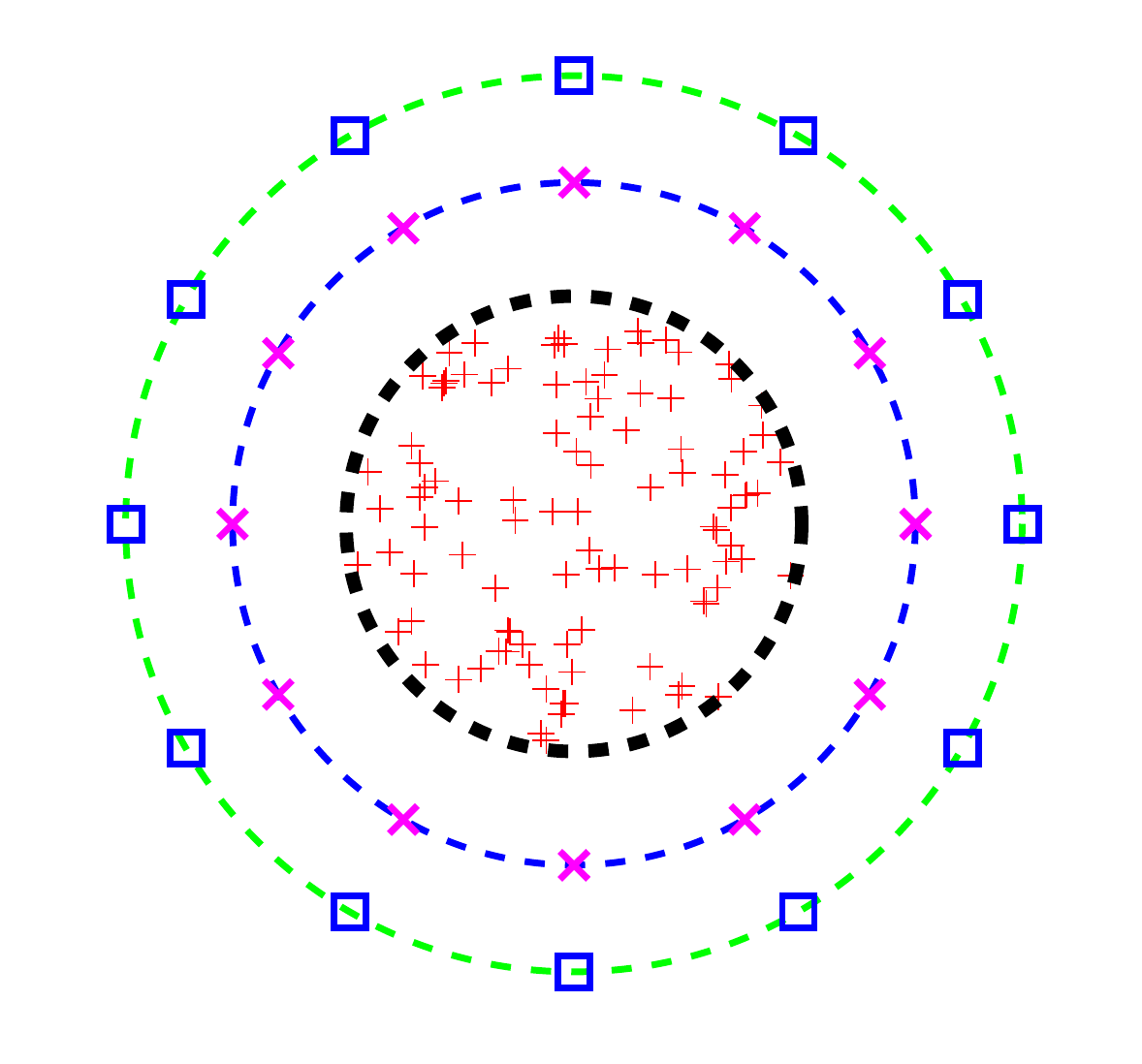}}
        \subfigure[\textbf{Algebraic.}\label{fig_skeletonization}]{\includegraphics[width=0.3\textwidth]{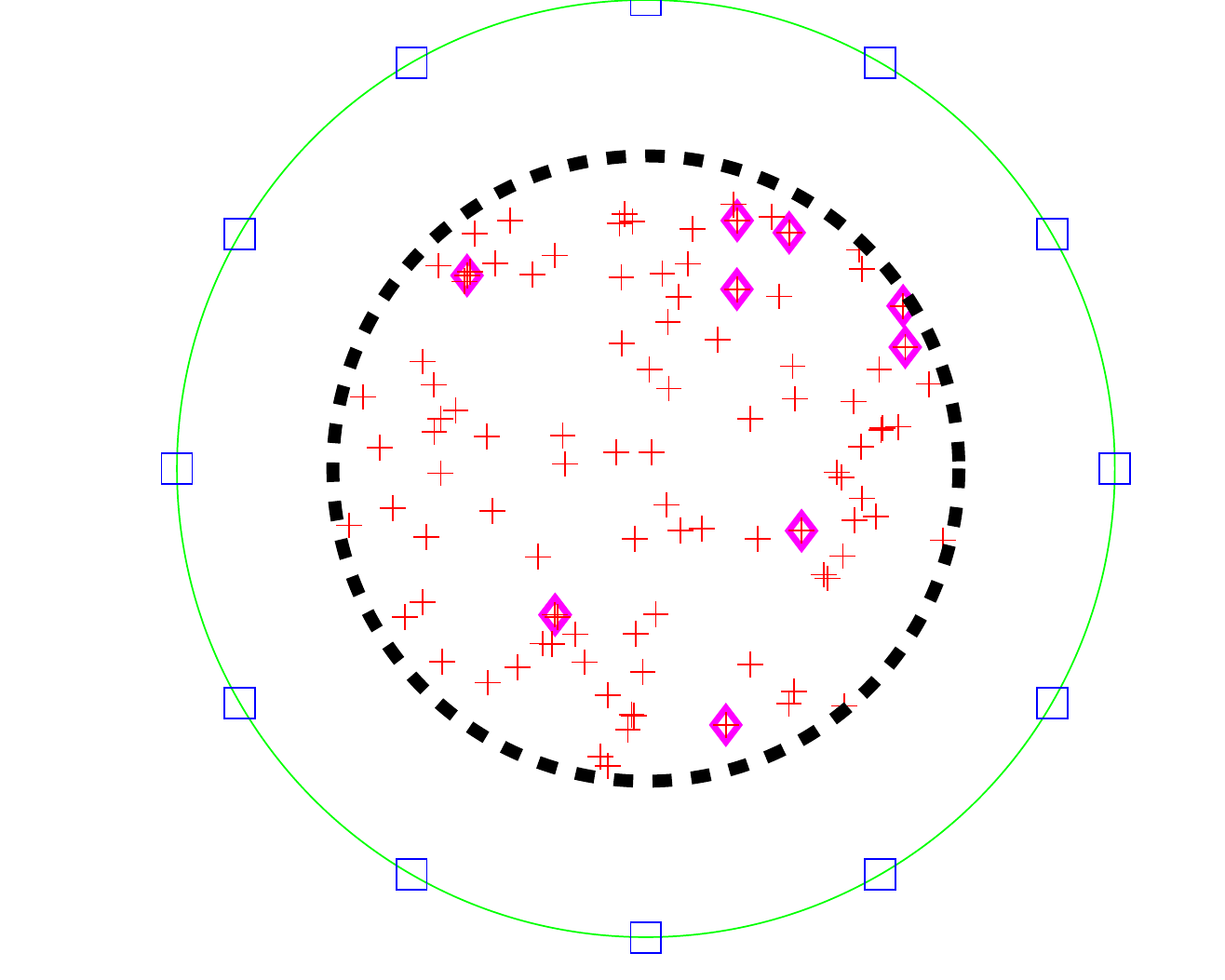}}
\caption{We illustrate three methods for computing an outgoing
        representation of the red source points. In
        Figure~\ref{fig_series}, we illustrate an analytic, single
        term expansion: the points are represented by their centroid.
        Higher order approximations can be viewed as Taylor expansions
        around this point and require a number of terms that grows significantly
        with the dimension $d$. In Figure~\ref{fig_interpolation}, we
        show a method based on placing equivalent sources and finding
        equivalent densities that can approximate the far
        field~\cite{ying-biros-zorin-03}.  An outgoing representation is
        constructed so that the far field due to the true sources (red
        points) is reproduced by equivalent sources (magenta ``X'').
		The charges on the equivalent sources are determined from interactions 
		with fictitious check points (blue squares). As
        the dimension of the input increases, the number of equivalent
        sources and check points required grows quickly, since they must cover 
		the surface of a bounding sphere or cube in $d$ dimensions. In
        Figure~\ref{fig_skeletonization}, we illustrate the
        skeletonization-based approach.  
        Using the interactions between the sources and fictitious targets (blue
        squares), the method computes
        an interpolative
        decomposition and chooses some skeleton sources (magenta points)
        to represent the far field. The number of skeleton points needed
        depends on the local intrinsic dimensionality and the
        kernel, rather than $d$. 
        However, with existing techniques, the number of targets
        needed can grow with the ambient dimension.
\label{fig_outgoing_rep_illustrations}}
\end{figure}

As we mentioned, the far-field approximation is much more complicated,
and there is a great variety of options which we discuss
in~\cite{march-biros14}. However, for completeness we summarize that
discussion in the Figure~\ref{fig_series}. In the left subfigure, we
depict analytic expansions based on series truncation
(e.g.,\cite{greengard-rokhlin-87}). These methods compute a series expansion
using only the source points. In general, the number of terms needed for a 
given accuracy scales unfavorably with $d$. The middle subfigure illustrates
the kernel independent fast multipole method which is a hybrid of
algebraic and analytic methods~\cite{ying-biros-zorin-03}. In this approach, 
the far-field is approximated via interactions between carefully chosen 
fictitious source and target points. These points are chosen to cover a bounding
sphere or box, so they also scale poorly with the ambient dimension. 
 The last figure shows a purely algebraic approach similar
to~\cite{martinsson-rokhlin07}.  Here, a subset of the source points is used 
in place of fictitious sources.  However, a number of fictitious targets which 
scales exponentially with $d$ is still necessary. 

The basic conclusion is that all of these existing
methods for constructing the far-field low-rank approximation do not
scale with increasing ambient dimension $d$. Furthermore
distance-based pruning also doesn't scale with increasing
dimensionality because even if the dataset has low intrinsic
dimension, the bounding box can be huge so that no pruning takes
place. 

\ASKIT{} introduces a new pruning method to approximate the far
field.  The effectiveness of both of these methods depends only on the
intrinsic dimension and not the ambient dimension.  In the remainder
of this section, we describe in detail how we carry out each of these
steps in \ASKIT.

\begin{figure}
  \centering
  \subfigure[\textbf{Fictitious targets.}\label{fig_fict_targets_id}]{\includegraphics[width=0.4\textwidth]{figures/algebraic_interpolation.pdf}}
\subfigure[\textbf{Subsampling (new).}\label{fig_subsampling_id}]{\includegraphics[width=0.4\textwidth]{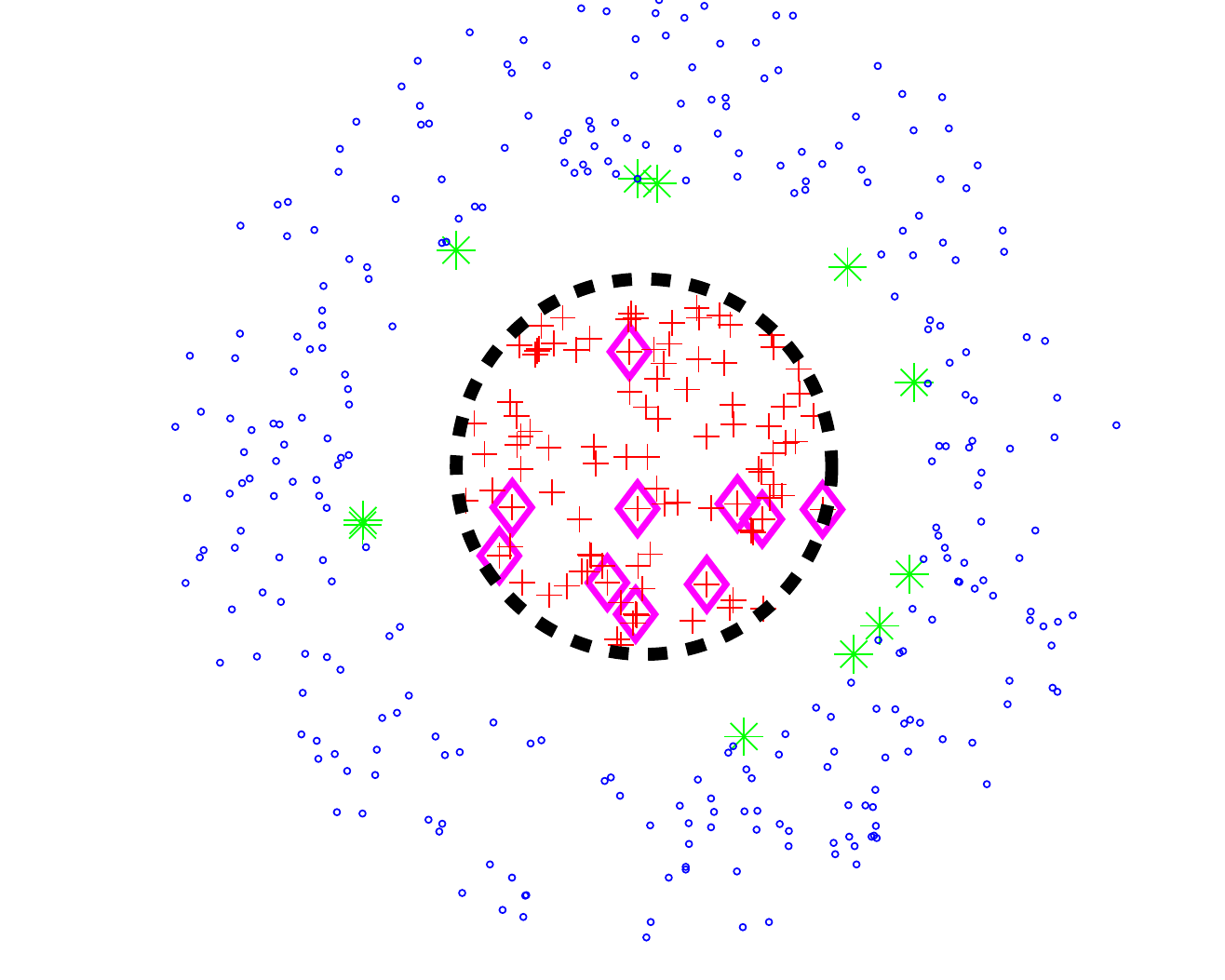}}
   \caption{\textbf{Two approaches for computing an algebraic outgoing
       representation.} In both cases, we are interested in computing
     an outgoing representation of the red source points. One method
     \cite{martinsson-rokhlin07} (Figure \ref{fig_fict_targets_id}),
     places a set of fictitious targets covering a ball or box
     surrounding the sources.  It computes the matrix of interactions
     between targets and sources, and computes its ID.  Our approach
     (Figure \ref{fig_subsampling_id}) subsamples $s$ of the
     well-separated target points (shown in green). We then compute
     the ID of the resulting $s \times n$ matrix $\subK$.
  \label{fig_algebraic_sampling}}
\end{figure}

\msubsection{Interpolative Decompositions and Sampling} The first
main component of our method is the representation of the far field
generated by source in a node using an {\em approximate} ID scheme,
which is summarized in Figure~\ref{fig_algebraic_sampling}.

Let $K \in \reals^{n \times m}$. Let $\MA{S}$ be an index set with
$\card{\MA{S}}=s$ and $1 \leq \MA{S}_j \leq m$. Let $K_S=K(:,\MA{S})$
be the columns of $K$ indexed by $\MA{S}$ and $K_R$ be the remaining
{\em unskeletonized} columns of $K$. Assuming $s<m<n$ and that $K_S$
is full rank, we can approximate the columns of $K_R$ by $K_S P$ where
$P=K_S^\dagger K_R, P\in \reals^{s\times m}$.  The ID consists of the
index set $\MA{S}$, referred to as the \emph{skeleton}, and matrix
$P$.  Following~\cite{
  martinsson-rokhlin-tygert07,cheng-gimbutas-martinsson-rokhlin05}, we
refer to the construction of this approximation for a matrix as
\emph{skeletonization}.

In order to compute an ID such that $\|K_R - K_S P\|$ is small, we
employ a pivoted QR factorization to obtain $K \Pi = QR$ for some
permutation $\Pi$, an orthonormal matrix $Q$, and upper triangular
matrix $R$.  The skeleton $\MA{S}$ corresponds to the first $s$
columns of $K\Pi$, and the matrix $P$ can be computed from $R$ in
$\bigO(s^3 + s^2(m-s))$ time.  It can be
shown~\cite{halko-martinsson-tropp11} that
\begin{equation}\label{e:id_error}
  \|K_R -K_S P\|\leq \sqrt{1 + m (m-s)} \sigma_{s+1}(K),
\end{equation}
where $\sigma_{s+1}$ is the $i^{\textrm{th}}$ singular value of
$K$. The overall cost for $s< m< n$ is $\bigO(n m^2)$. 

{\bf Far field using skeletonization.} We will use ID to compactly
represent the far field of a leaf node $\alpha$. Let $\XXa$ be the set
of points assigned to $\alpha$ (assume $\card{\XXa}=\ppl$). Let
$\bKa:=K(\XX\sdf\XXa,\XXa) \in \reals^{(N-\ppl) \times \ppl}$.  Also
let $w_\alpha=w(\XXa) \in \reals^\ppl$.  Our task is to construct an
approximation to $\bKa w_\alpha$.  

We choose $\ns$, compute the skeleton $\Sa$ of $\bKa$ and set
\begin{equation}
\bKa w_\alpha \approx \bKas
\tilde{w}_\alpha,\ \mbox{where}\  \bKas:=\bKa(:,\Sa),\ \tilde{w}_\alpha:= w_\alpha(\Sa) +
P_\alpha w_\alpha(\MA{R}),
\end{equation} 
and $\MA{R}$ is the index set of the unskeletonized columns of
$\bKa$. We term $\tilde{w}_\alpha \in \reals^\ns$ the {\em skeleton
  weights}.

Given the skeleton $\Sa$ and skeleton weights $\tilde{w}_\alpha$, we
can efficiently approximate the contribution to some target point ${\b
  u_i}$. We first compute the $1 \times s$ matrix of interactions
$K({\b u},\Sa)$, then apply it to the vector $\tilde{w}_\alpha$, to
get an approximation with error bounded by \eqref{e:id_error}.

This approach leaves two issues unanswered: first, how do we choose
$\ns$? We discuss this in Section~\ref{s:complexity-and-error}. Second,
computing the ID as described is more expensive than directly evaluating $\bKa
w_\alpha$. We address this point next.

\textbf{Approximate skeletonization.}  We have $\bigO(N/\ppl)$ leaves
and the skeletonization of each leaf described above 
costs $\bigO(N \ppl^2)$. When
performed for all leaf nodes, this will require $\bigO(N^2\ppl)$ work.
Instead, we will compute the skeleton of a smaller matrix, which has
only a random subset of the rows of $\bKa$.  That is, we select
$\ssize$ rows with $\ppl<\ssize\ll N$ and whose index set we denote by
$\XTa$. We form $\bKa(\XTa,:) \in \reals^{\ssize \times \ppl}$, and
compute the skeleton $\Sa$ of size $\ns$ and the corresponding
projection matrix $P_\alpha$.  This is equivalent to choosing $\ssize$
target points. The complexity of the construction for one leaf becomes
$\bigO(\ssize \ppl^2)$; thus the overall complexity for all nodes in
the tree becomes $\bigO(N\ssize \ppl)$. This approach is illustrated in 
Figure~\ref{fig_algebraic_sampling}.

\textbf{Sampling rows of $K$.} We need to choose a small number of
rows such that the ID of $\bKa(\XTa,:)$ will be close to the ID of
$K$. Randomized linear algebra algorithms can achieve this by either
random projections \cite{halko-martinsson-tropp11} or the construction
of an importance sampling distribution \cite{mahoney-drineas09}.
Either approach requires $\bigO(N)$ work per node.  However, for
smoother kernels that decay with distance (or, more generally,
dissimilarity) the nearest (more similar) points will tend to dominate
the sum in~\eqref{e:near-far}. Following this intuition, if we can
include the nearest neighbors of each point in $\XXa$, then we expect
this to be a reasonable approximation to an importance sampling
distribution. 
 If we do not have enough neighbors, we add additional
uniformly chosen points to reach a sufficient sample size. This
process is discussed below.
\begin{figure}[tbph]
	\centering
        \includegraphics[width=0.45\textwidth]{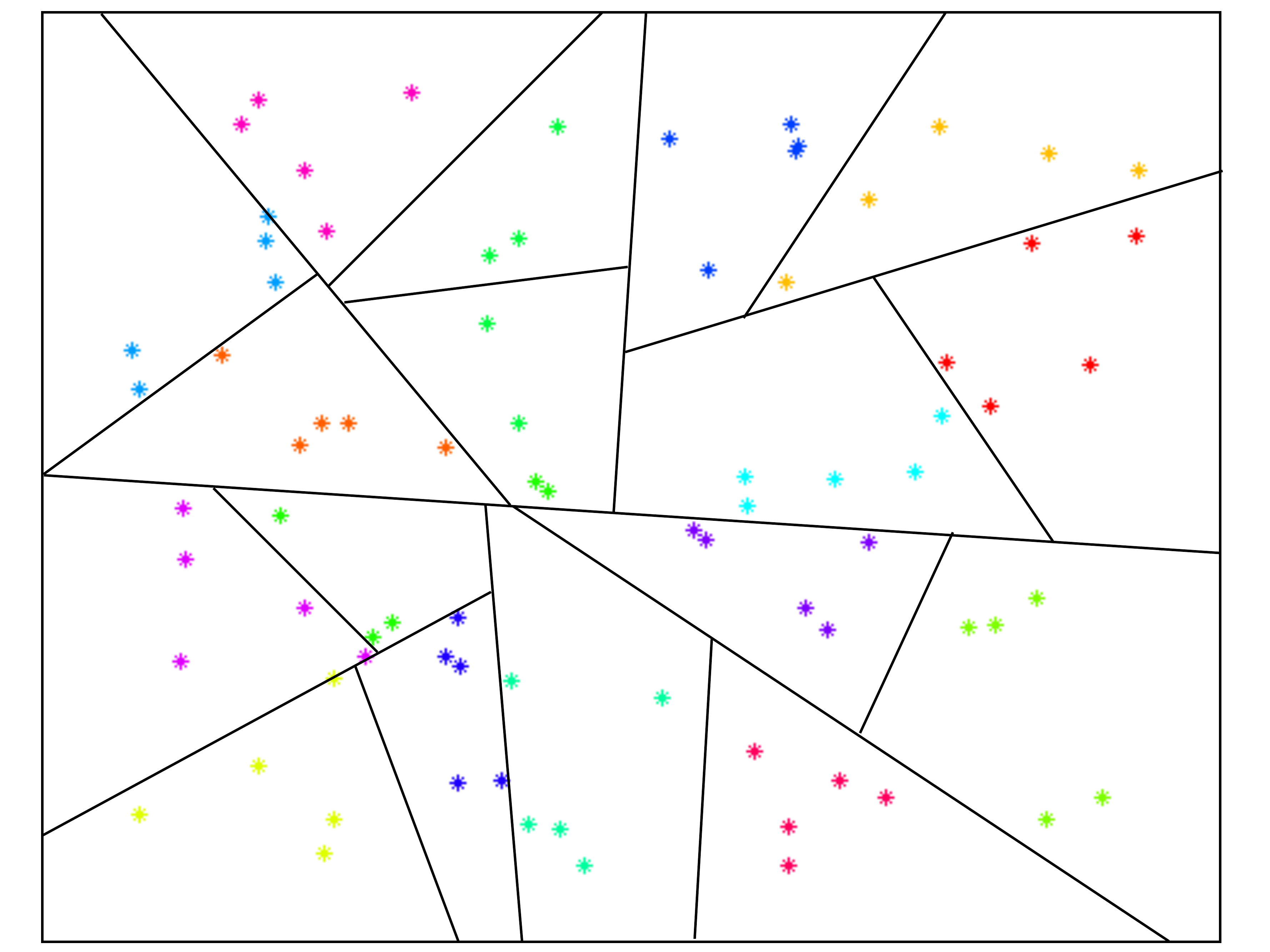}
        \caption{\textbf{The tree.}  An illustration of our
          median-split binary tree. We split a node by estimating the
          farthest pair of points in it.  We project all points onto
          the line between these points, then split at the median.
          The splitting planes are shown, and points belonging to
          different nodes are shown in different colors.
        \label{fig_tree}}
\end{figure}
\begin{figure}[tbph]
	\centering
        \subfigure[A leaf node.\label{subfig_leaf_1}]
        		{\includegraphics[width=0.45\textwidth]{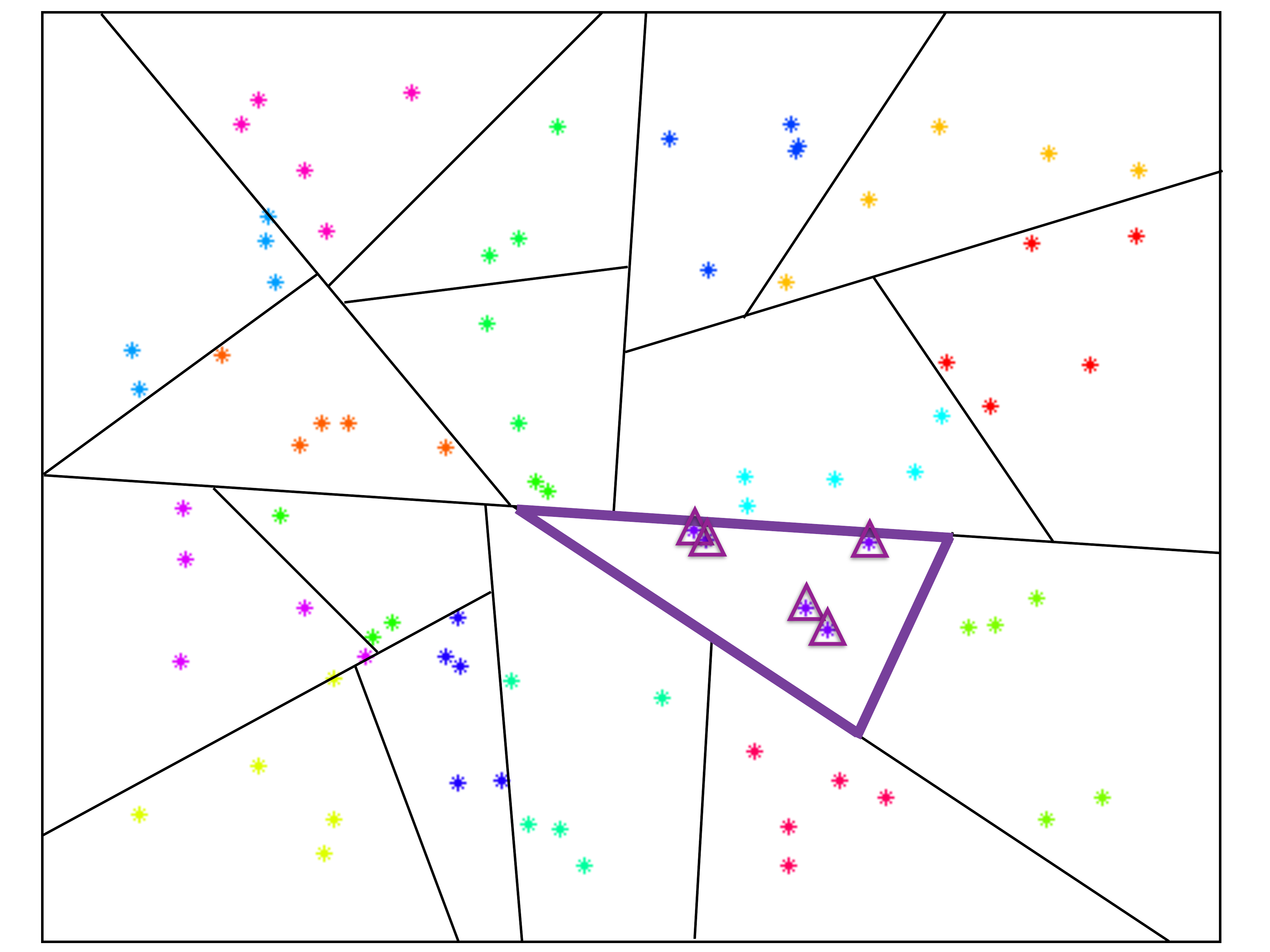}}
        \subfigure[The node's nearest neighbors.\label{subfig_leaf_2}]
		{\includegraphics[width=0.45\textwidth]{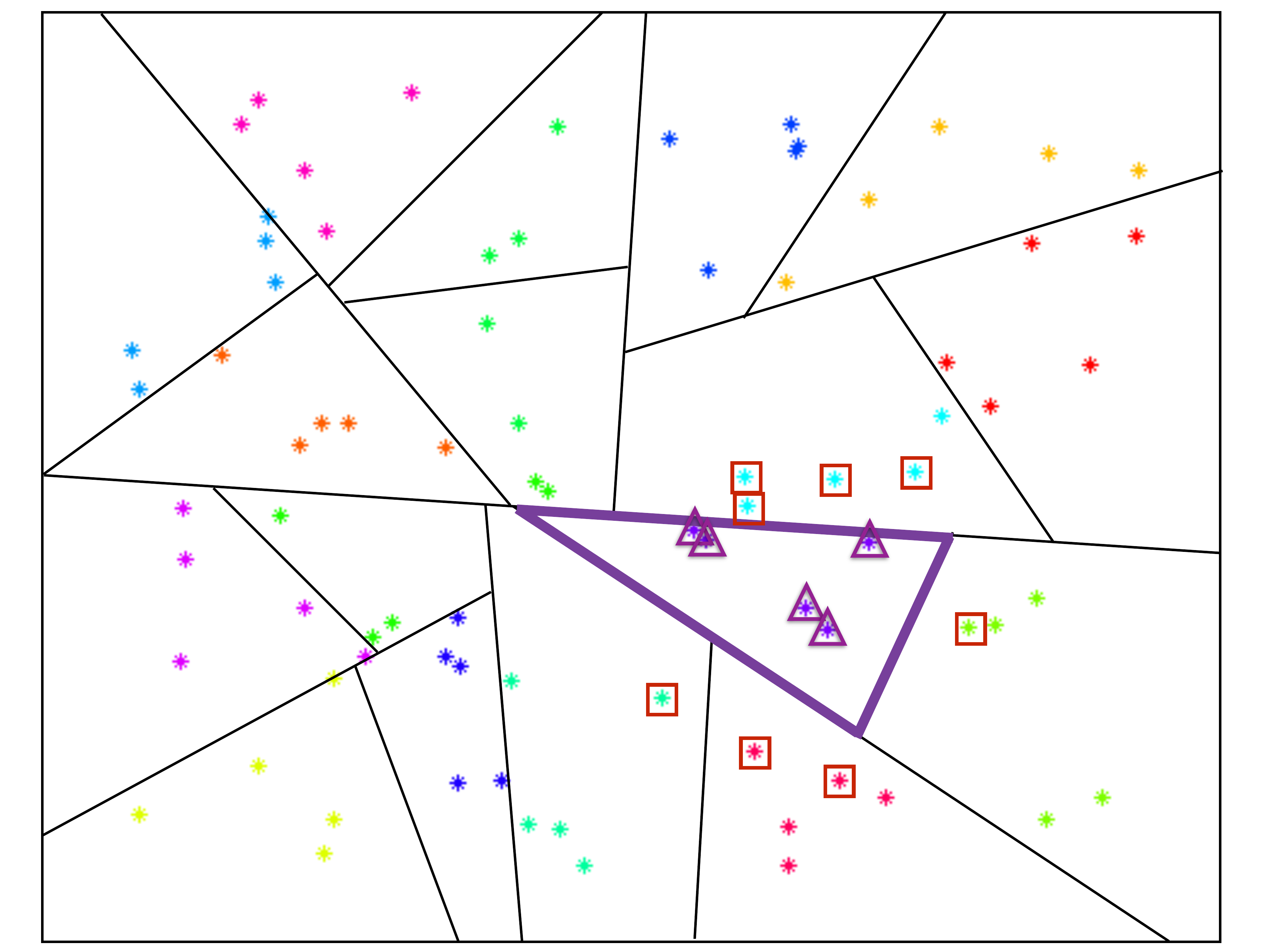}}
        \subfigure[Sampling distant points.\label{subfig_leaf_3}]
		{\includegraphics[width=0.45\textwidth]{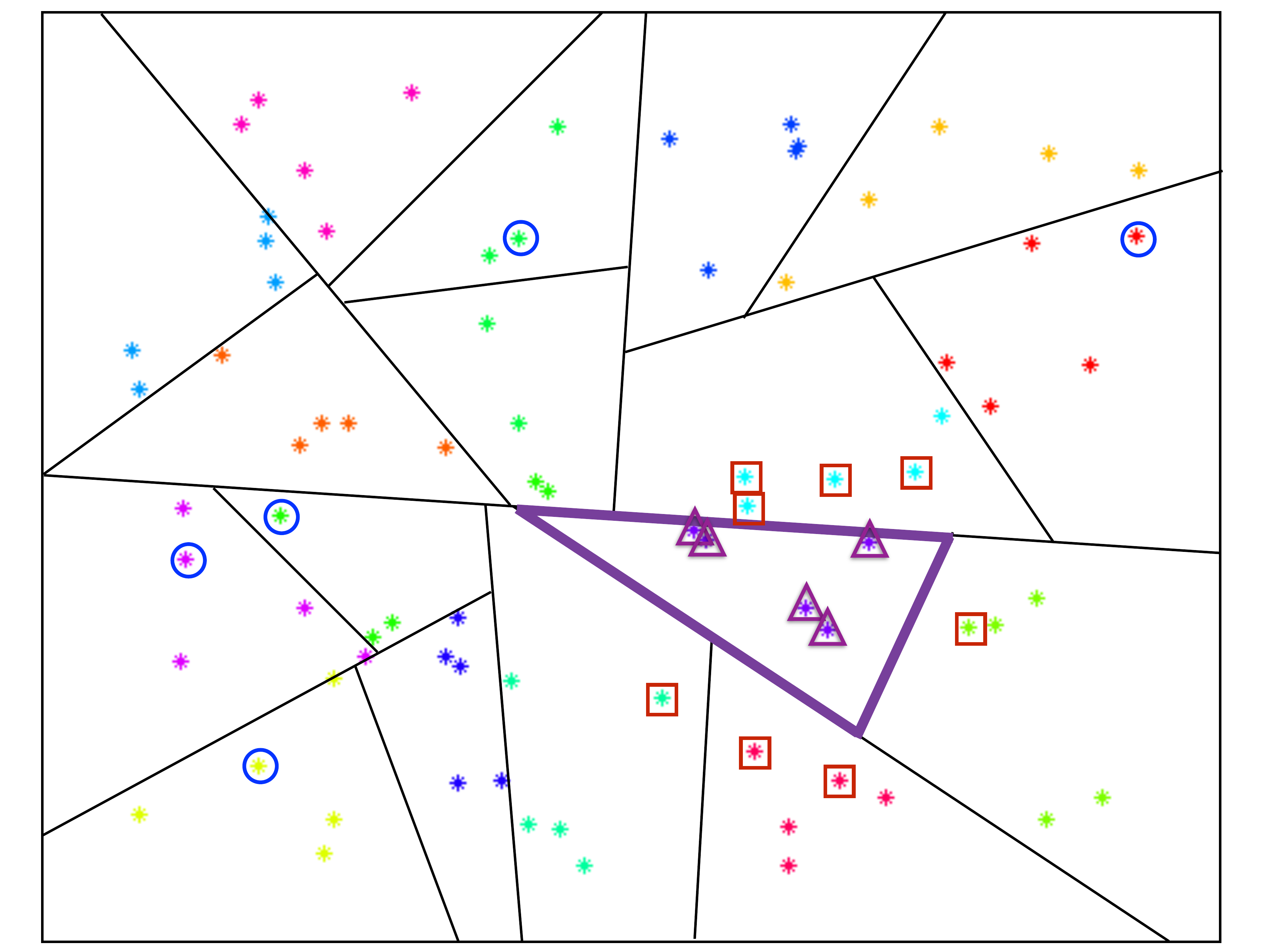}}
        \subfigure[The resulting skeleton.\label{subfig_leaf_4}]
        {\includegraphics[width=0.45\textwidth]{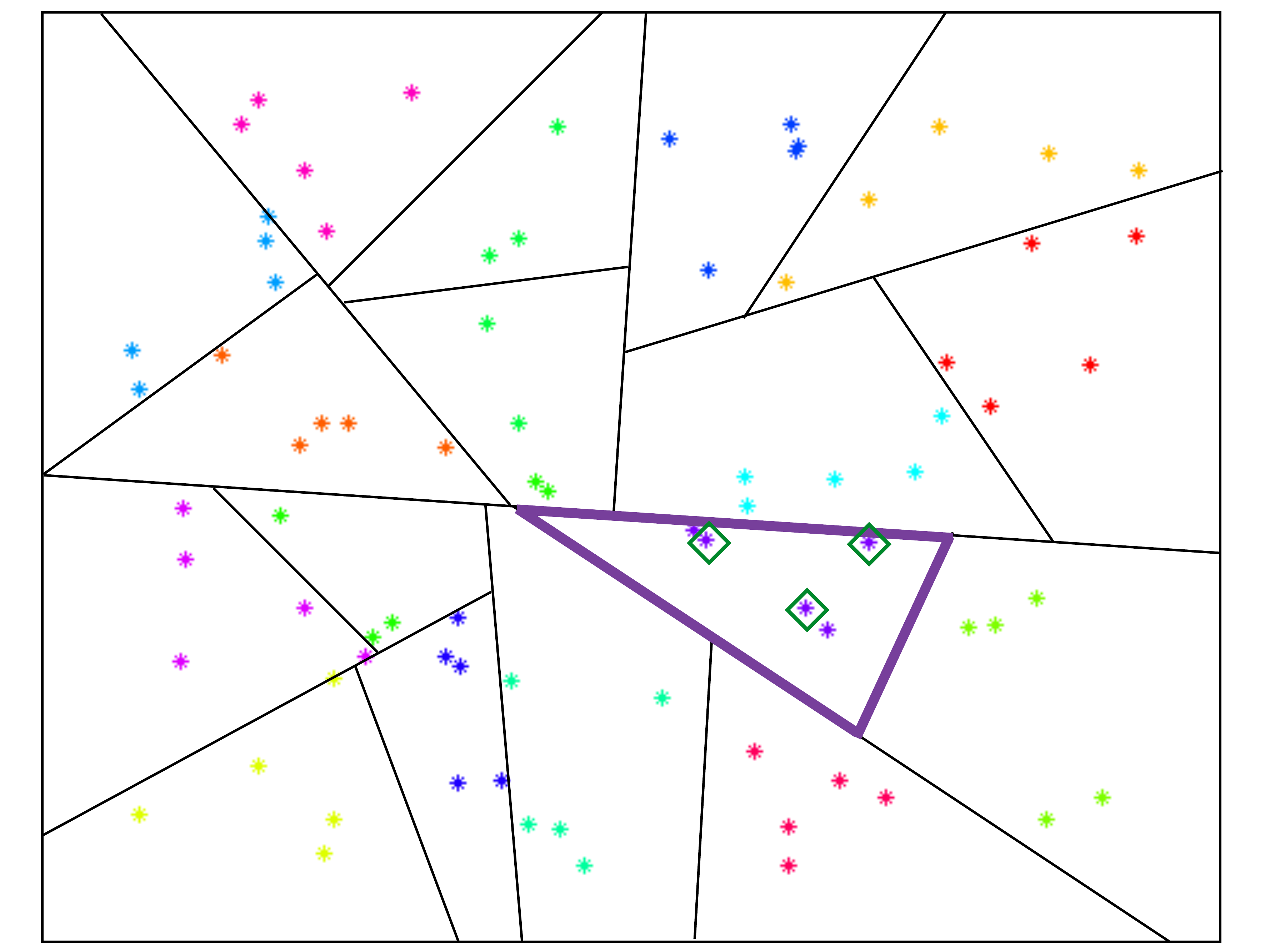}}
	\caption{\textbf{Skeletonizing a leaf.}
    \textbf{Fig.~\ref{subfig_leaf_1}.} We highlight the leaf node to
    be skeletonized. The points to be approximated are shown with
    triangles.  \textbf{Fig.~\ref{subfig_leaf_2}.} We compute the
    union of the lists of nearest neighbors of the points in the leaf,
    and exclude points that belong to the leaf itself.  These points
    are highlighted with squares.  \textbf{Fig.~\ref{subfig_leaf_3}.}
    We sample additional distant points for the skeletonization.
    These points are highlighted with circles. We compute the matrix
    of interactions with rows given by the neighbors and samples
    (squares and circles) and columns given by the points in the leaf
    (triangles).  \textbf{Fig.~\ref{subfig_leaf_4}.} We compute the ID
    of this matrix to obtain the skeleton points, highlighted with
    diamonds. Notice that the node has two more points that are not
    part of the skeleton and will not be used for far-field
    evaluations.  We can now approximate the contribution of the
    points in this node to a distant target point using only the
    interactions with these skeleton points.
	\label{fig_skeletonize_leaf}}
\end{figure}

\begin{figure}[tbph]
        \centering
        \subfigure[An internal node. \label{subfig_internal_1}]{\includegraphics[width=0.45\textwidth]{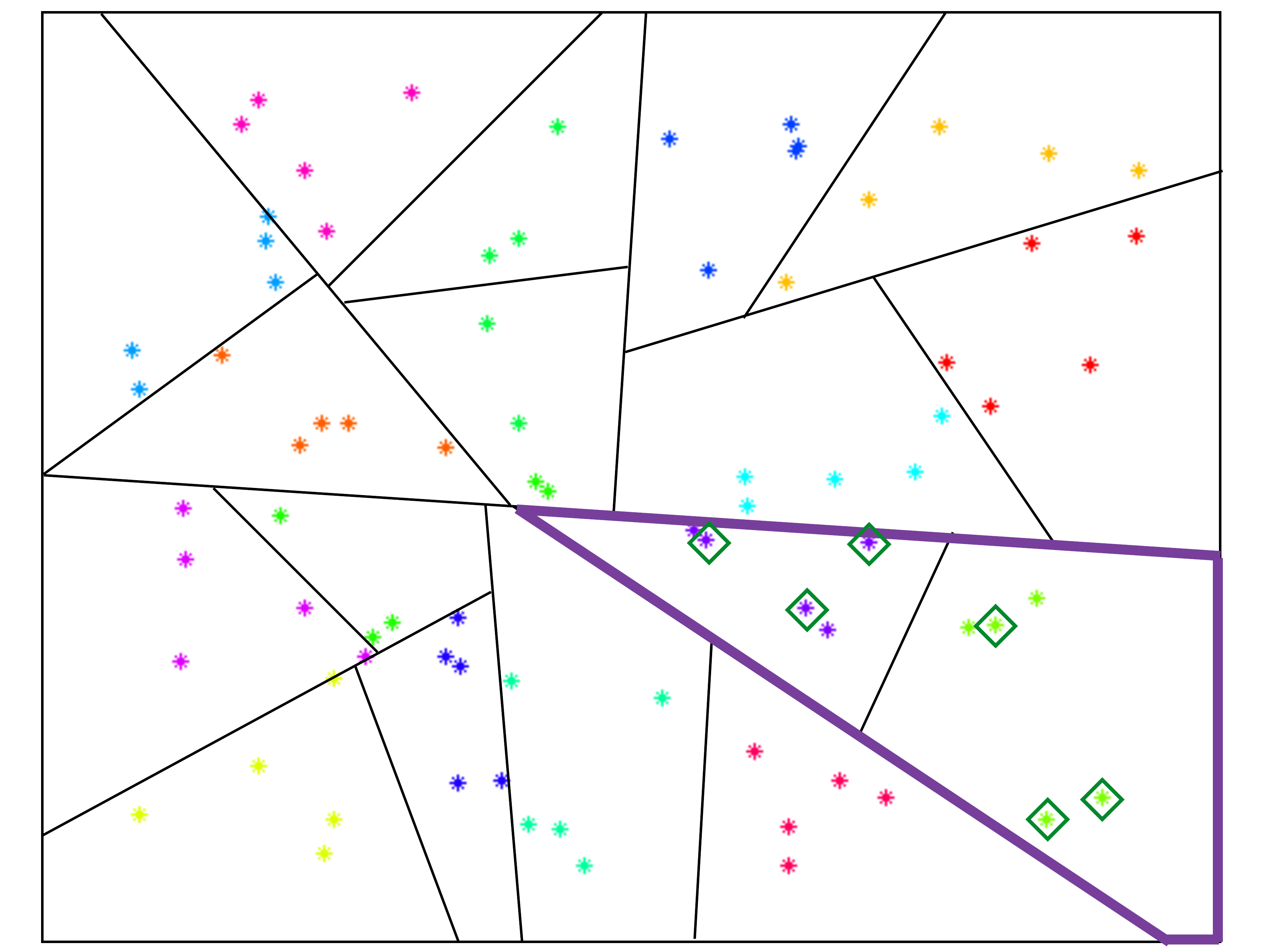}}
        \subfigure[Merging the neighbor lists. \label{subfig_internal_2}]{\includegraphics[width=0.45\textwidth]{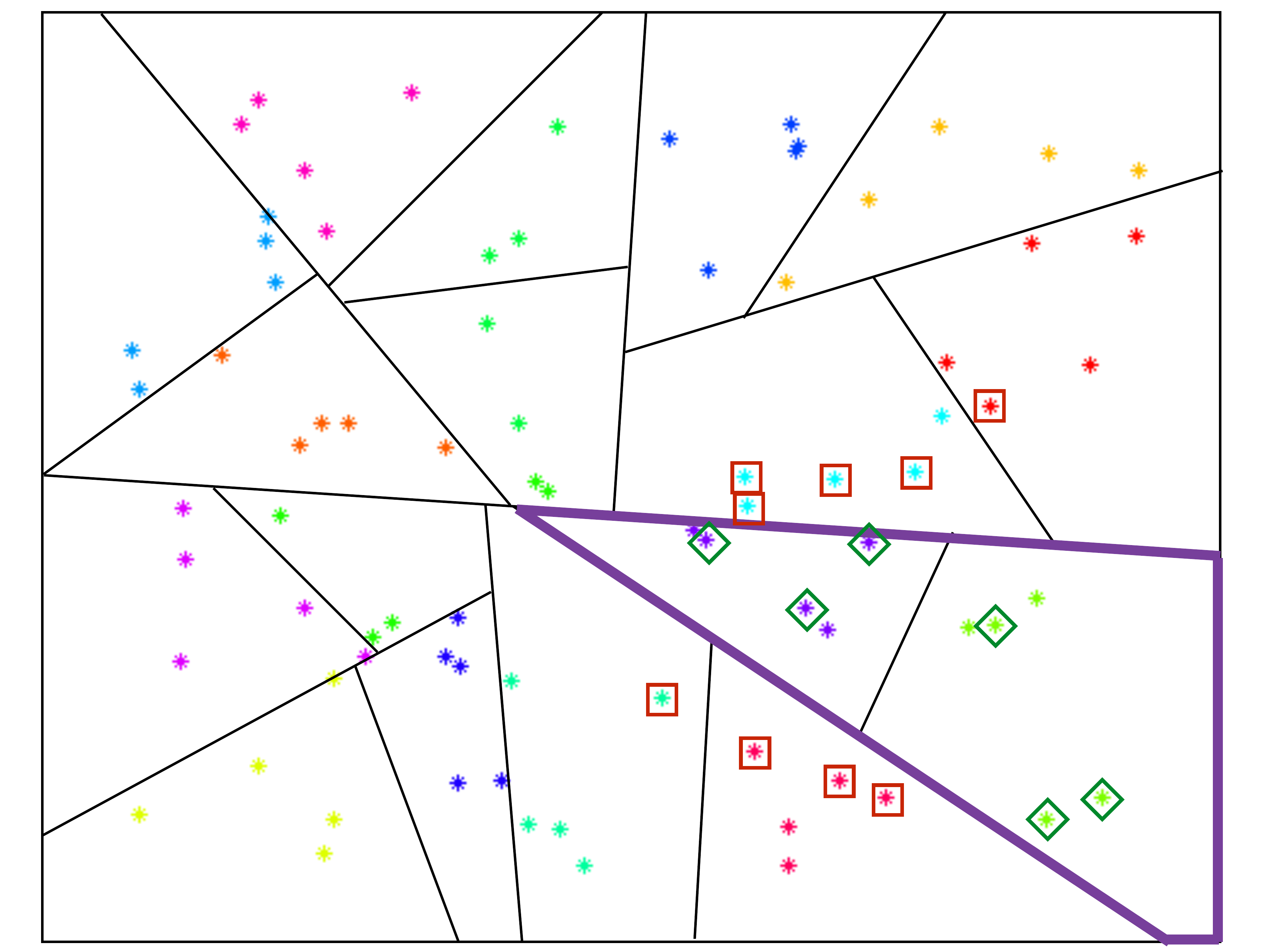}}
        \subfigure[Sampling distant points.\label{subfig_internal_3}]{\includegraphics[width=0.45\textwidth]{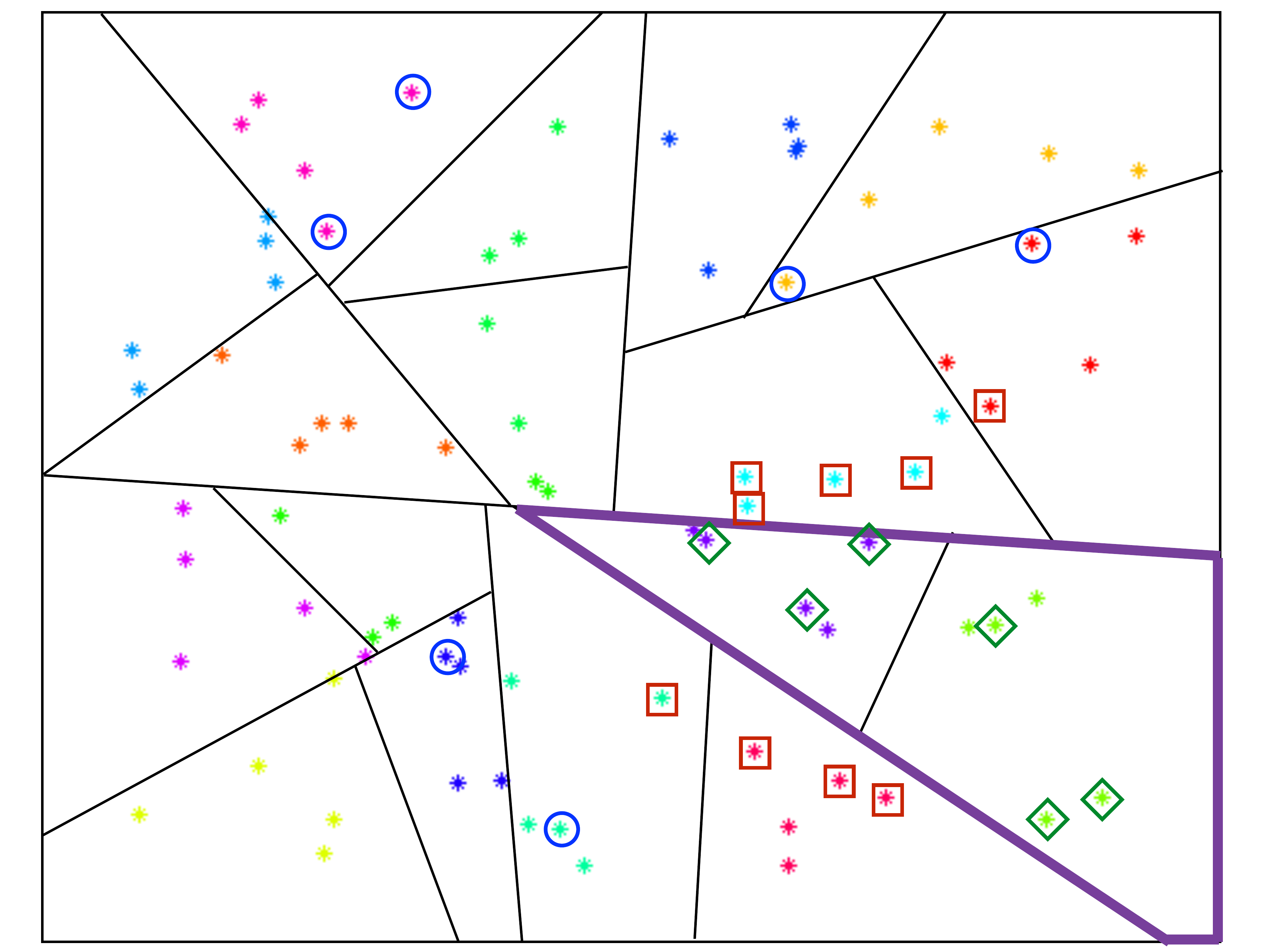}}
        \subfigure[The resulting skeleton.\label{subfig_internal_4}]{\includegraphics[width=0.45\textwidth]{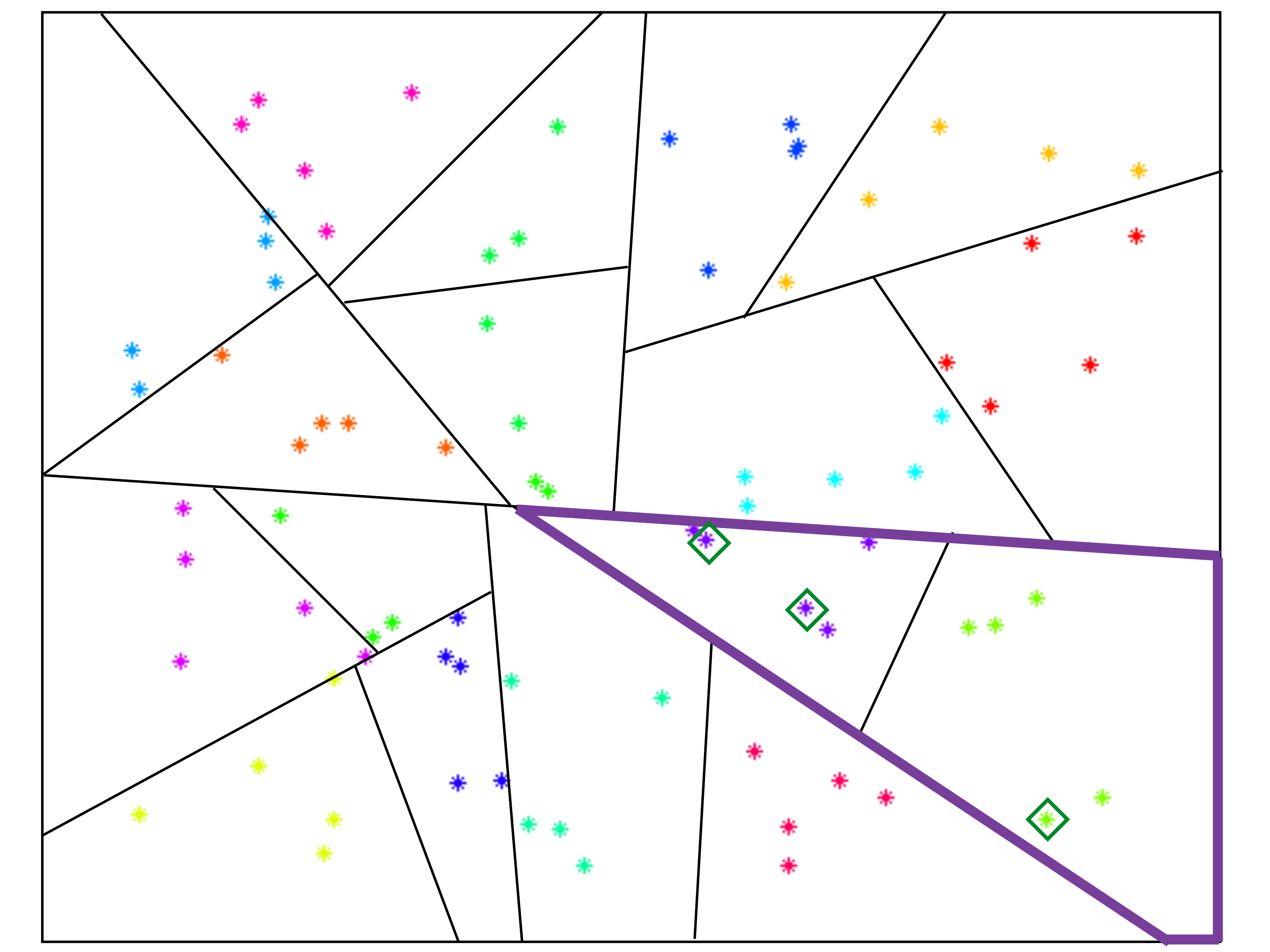}}
\caption{\textbf{Skeletonizing an internal node.}
  \textbf{Fig.~\ref{subfig_leaf_1}.} We skeletonize the highlighted
  internal node. The skeletons of its children are highlighted with
  diamonds.  \textbf{Fig.~\ref{subfig_leaf_2}.} We merge the nearest
  neighbor lists of the children, then exclude all of the points
  belonging to the node to be skeletonized. Finding these points can
  be done in $\bigO(\log (N/m))$ time using the Morton IDs.  These points are
  highlighted with squares.  \textbf{Fig.~\ref{subfig_leaf_3}.} We
  sample additional distant points, highlighted with circles.  We
  compute the matrix of interactions with rows given by the neighbors
  and samples (squares and circles) and columns given by skeletons of
  the child nodes (triangles).  \textbf{Fig.~\ref{subfig_leaf_4}.} We
  compute the ID of this matrix to obtain a skeleton for the parent
  node, which is a subset of the combined skeletons of the children.
\label{fig_skeletonize_internal_node}}
\end{figure}

\begin{figure}[tbph]
	\centering
        \subfigure[Evaluating at a target point (star).\label{subfig_downward_1}]{\includegraphics[width=0.45\textwidth]{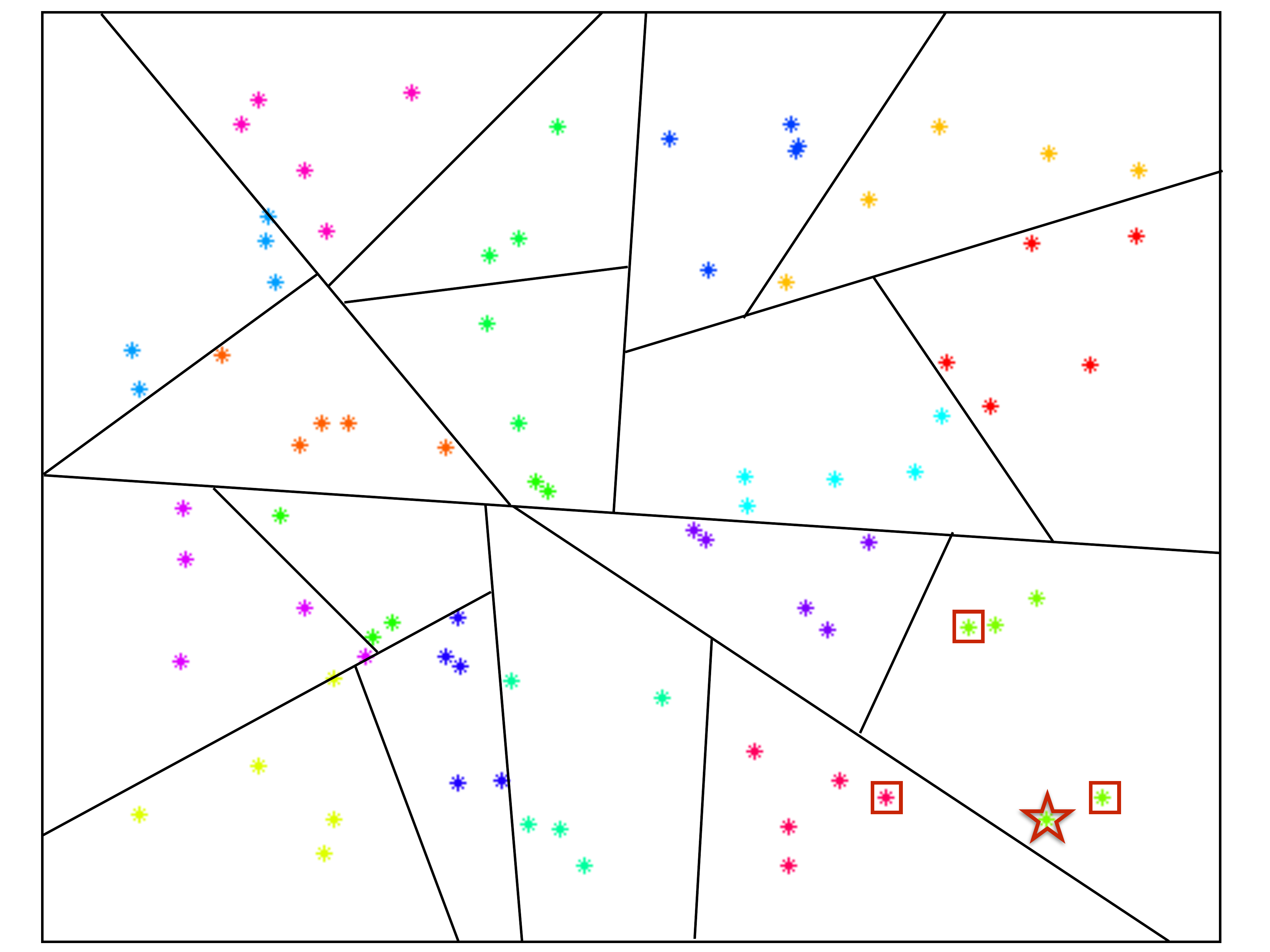}}
        \subfigure[Pruning the highlighted node.\label{subfig_downward_2}]{\includegraphics[width=0.45\textwidth]{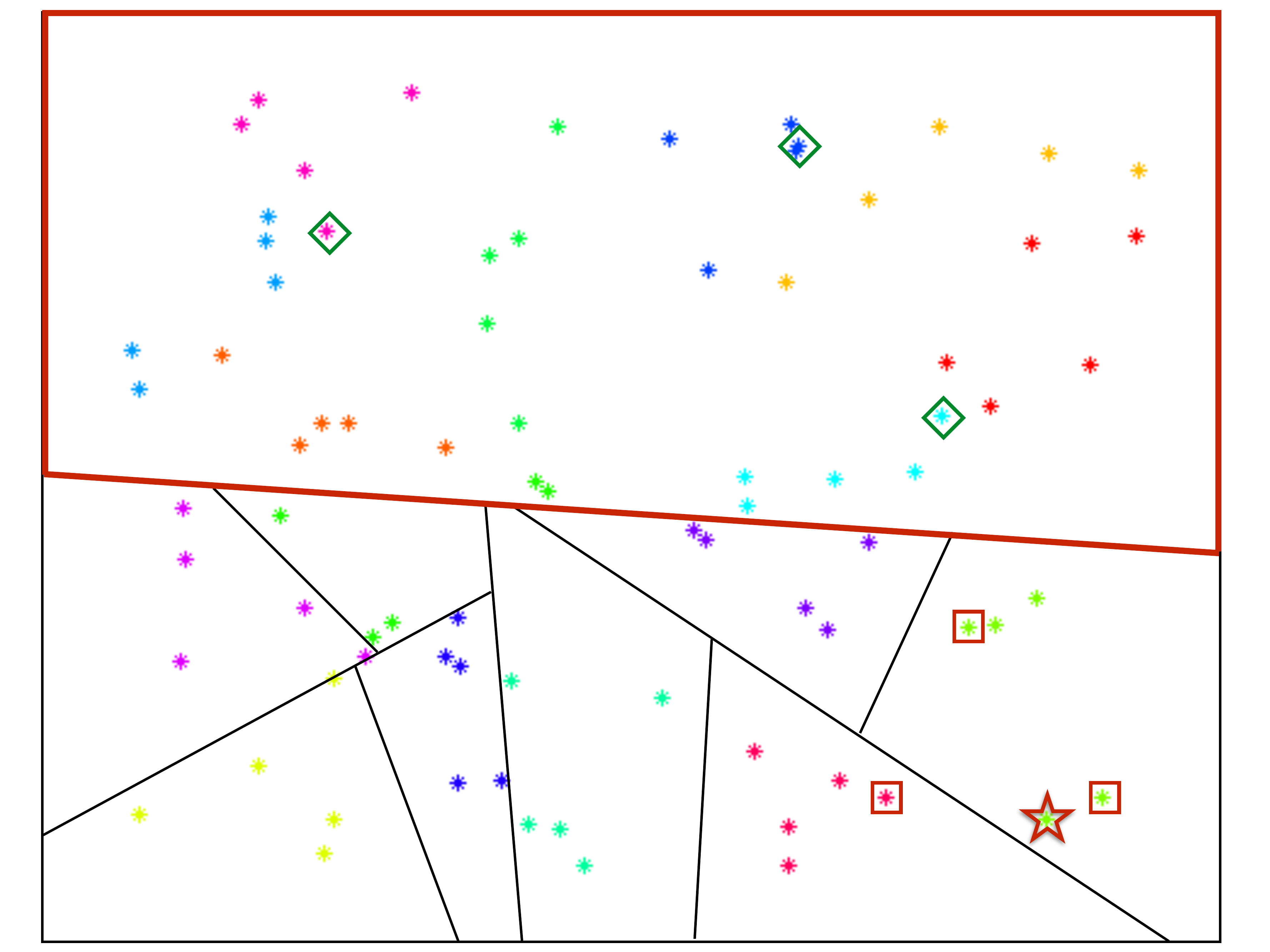}}
        \subfigure[Pruning another node.\label{subfig_downward_3}]{\includegraphics[width=0.45\textwidth]{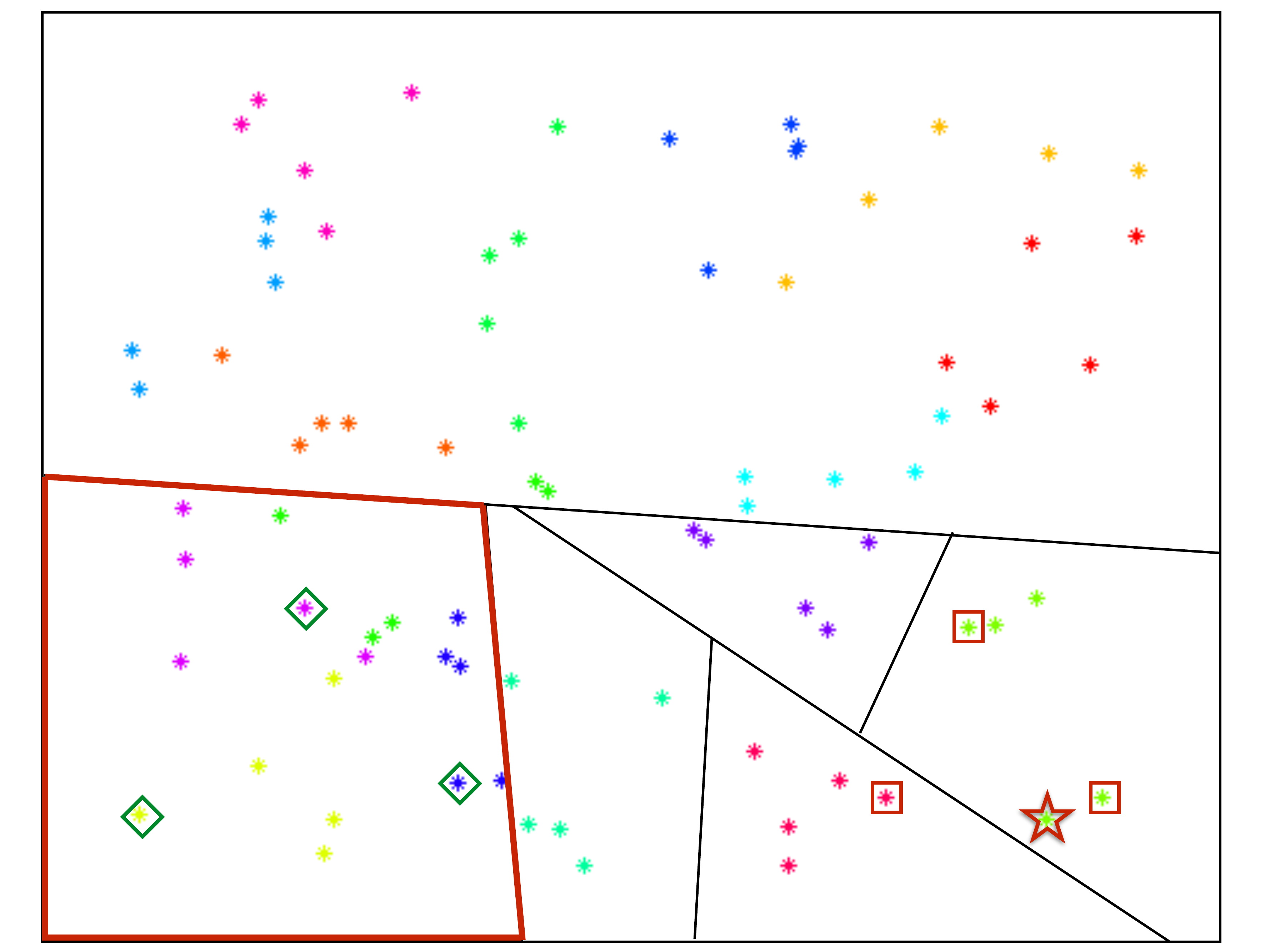}}
        \subfigure[Pruning a leaf node.\label{subfig_downward_4}]{\includegraphics[width=0.45\textwidth]{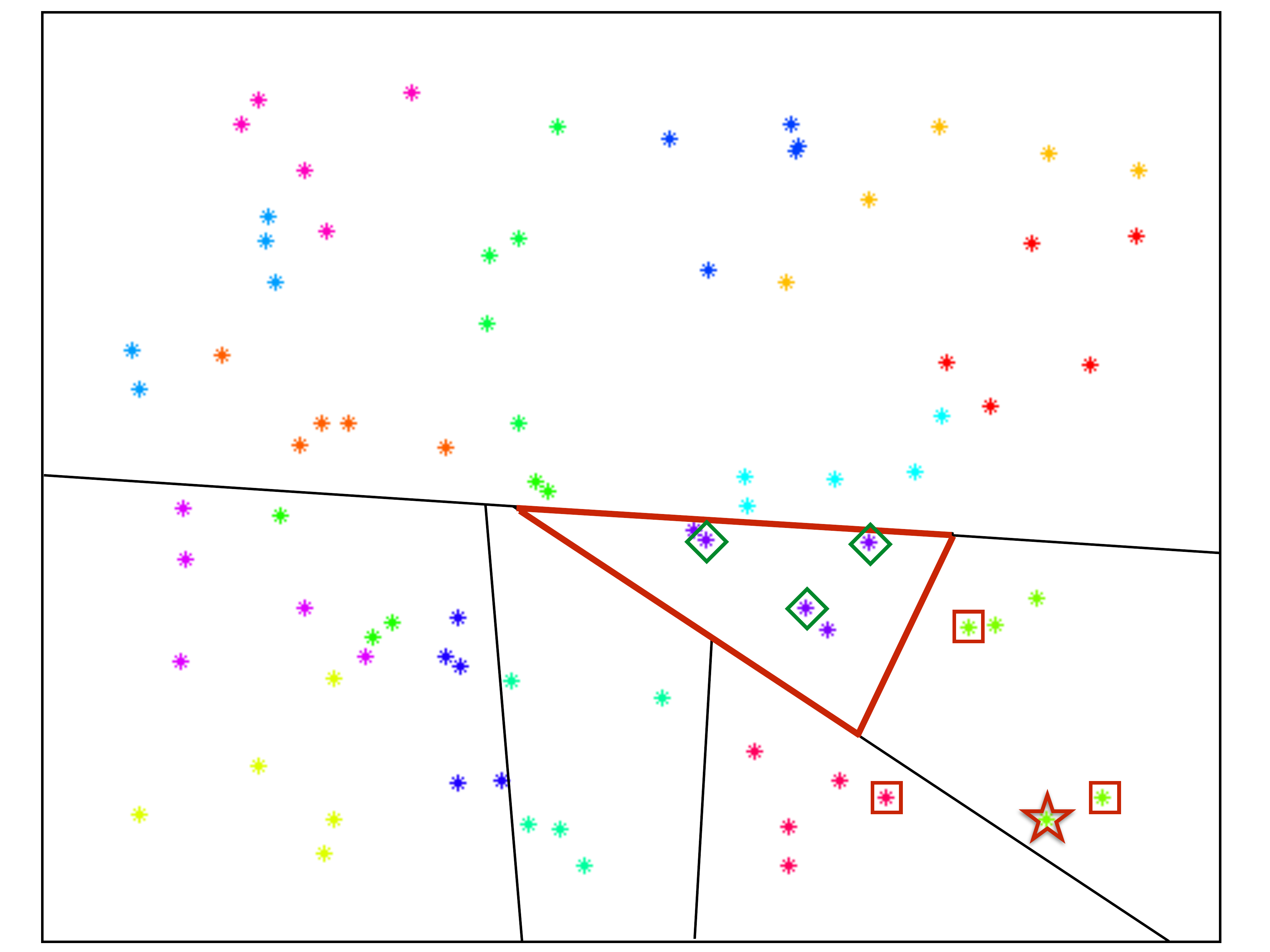}}
        \subfigure[Pruning another leaf node.\label{subfig_downward_5}]{\includegraphics[width=0.45\textwidth]{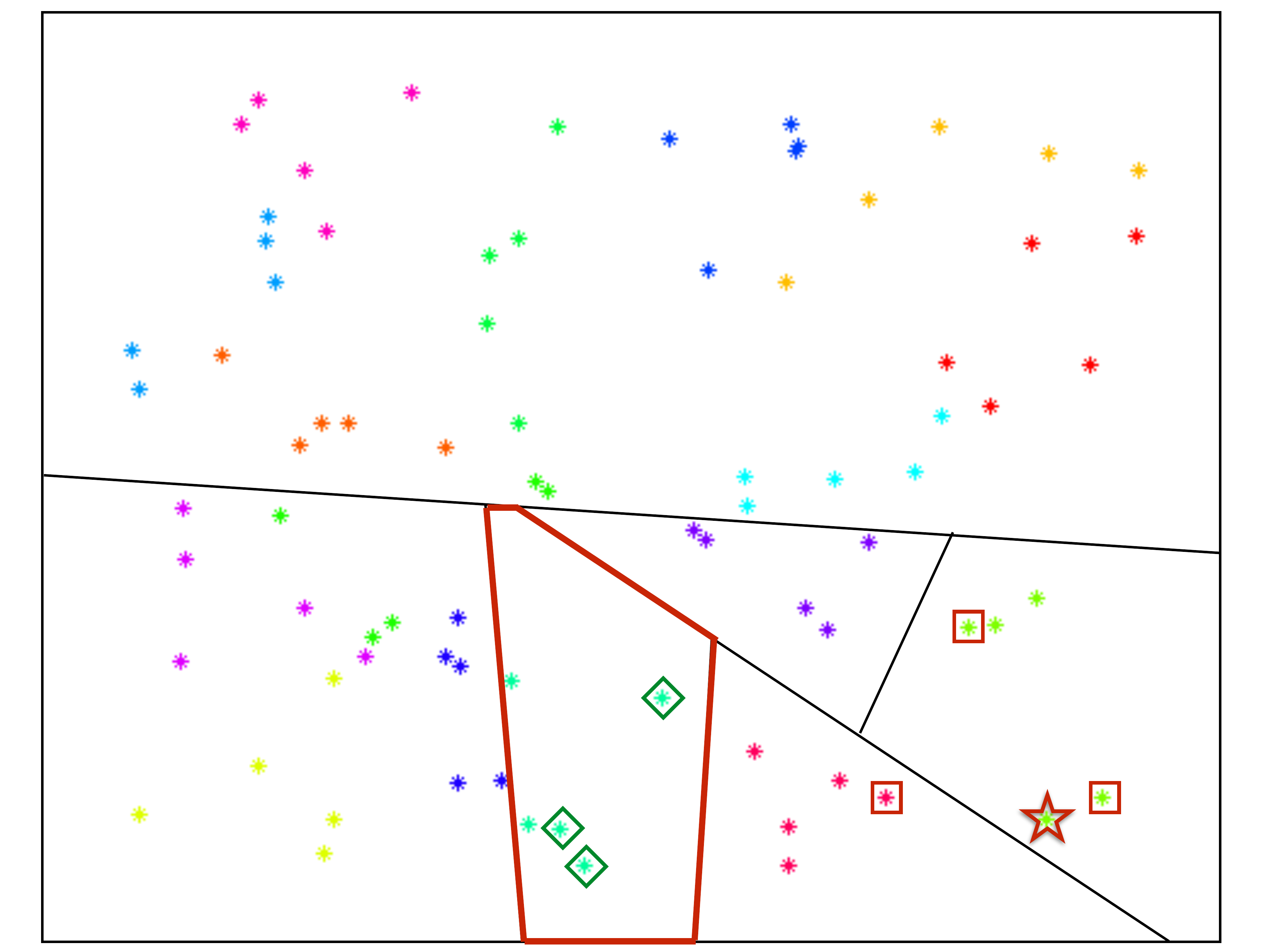}}
        \subfigure[Directly evaluating un-prunable leaves.\label{subfig_downward_6}]{\includegraphics[width=0.45\textwidth]{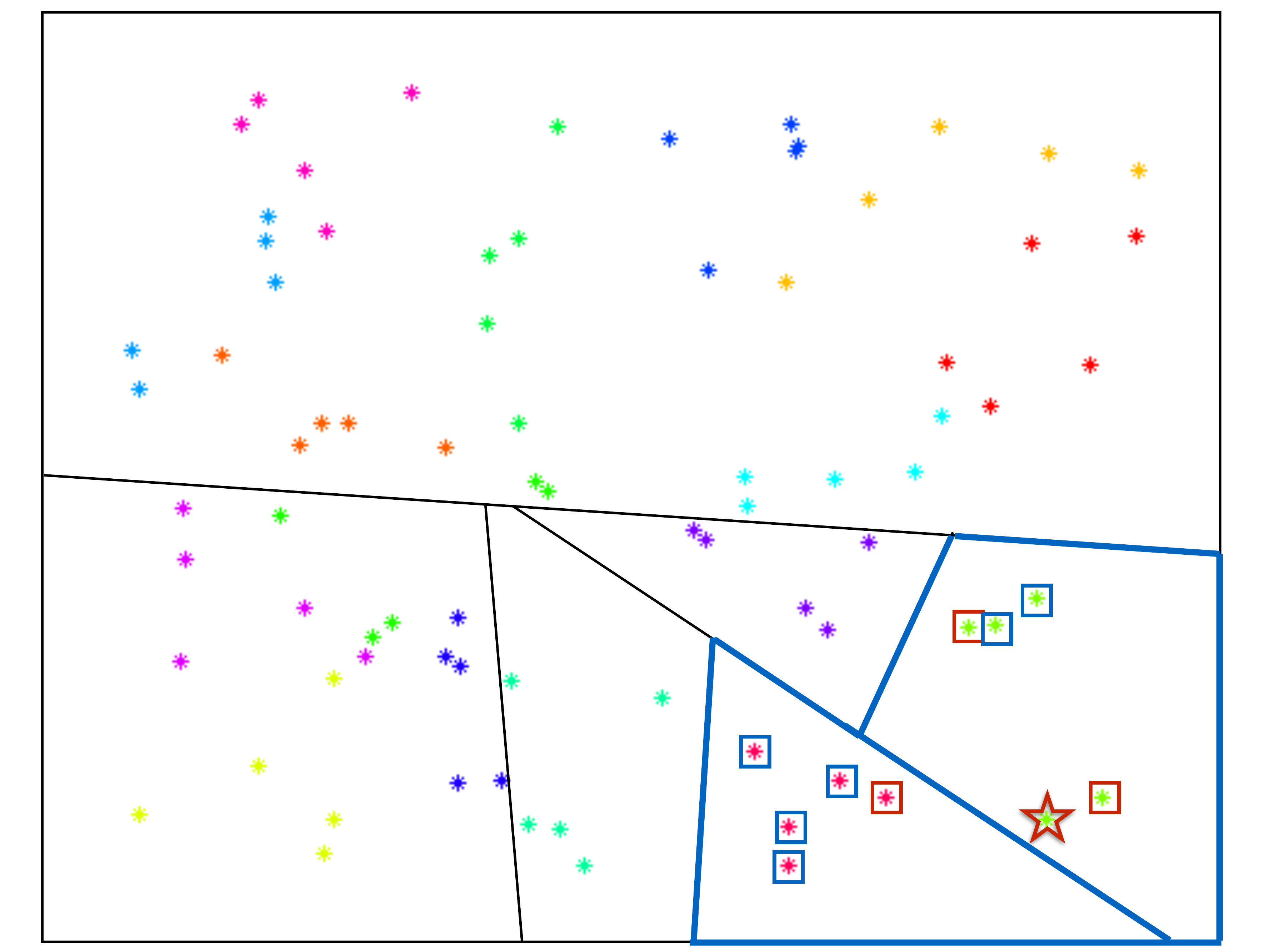}}
\caption{\textbf{Evaluating the approximate summation.}
  \textbf{Fig.~\ref{subfig_downward_1}.} We evaluate the potential at
  a target point, highlighted with a star. We show its nearest nearest
  neighbors as well, highlighted with squares.
  \textbf{Fig.~\ref{subfig_downward_2}.} We traverse the tree in a top
  down fashion.  We can prune a node if the node does not contain any
  neighbors of the target point. By pruning we mean evaluating the
  field of the node using the skeleton on the target point and then
  terminating the traversal.  The node highlighted in red satisfies
  our pruning criterion, so we evaluate its contribution
  approximately.  We compute the interactions between the skeleton
  points (diamonds) and the target. We use the skeleton weights for
  these points to compute the effective contribution of the node.
  \textbf{Fig.~\ref{subfig_downward_3}.} We continue traversing the
  tree.  Once again, we have a node that contains no nearest neighbors
  of the target, so we use the skeleton points to compute its
  approximate contribution.  \textbf{Fig.~\ref{subfig_downward_4}.} We
  still use the approximate representation at leaves which satisfy the
  pruning criterion.  \textbf{Fig.~\ref{subfig_downward_5}.} We
  continue the approximate evaluation of all nodes that satisfy the
  pruning criterion.  \textbf{Fig.~\ref{subfig_downward_6}.} The
  leaves highlighted in blue do not satisfy the pruning criterion, so
  we compute their contribution directly.  We evaluate the direct
  interaction between all points in these nodes (highlighted with
  squares) and the target point.
\label{fig_alg_steps}}
\end{figure}

\msubsection{\ASKIT\label{askit_subsec}} 
Using the ID as a compact representation, we can now describe the main
steps of our treecode:
\begin{itemize}
\item Approximate the $\kappa$-{\bf nearest neighbors} for all $\bx_i$.
\item Compute a top-down  {\bf binary tree decomposition} for $\XX$.
\item Perform a bottom-up traversal to build {\bf neighbor lists} for interior 
(non-leaf) nodes.
\item Perform a bottom-up traversal to compute {\bf skeletons and equivalent 
weights}.
\item Perform a top-down traversal to {\bf evaluate} $u_i$ at each point $i$.
\end{itemize}
We describe the individual steps below in detail. 

The basic steps of the algorithm are illustrated in
Figure~\ref{fig_tree} (leaves of a tree),
Figure~\ref{fig_skeletonize_leaf} (skeletonization of a leaf),
\ref{fig_skeletonize_internal_node} (skeletonization of an internal node) 
and \ref{fig_alg_steps} (evaluation).  

\textbf{Computing nearest neighbors.} To find
nearest neighbors we use a greedy search using random projection
trees~\cite{dasgupta-freund08}. We build a tree and for each $\bx_i$
we collect $\kappa$-nearest neighbors found by exhaustive search among the
other points in the leaf node that contains $\bx_i$. Then we discard
the tree (we do not perform top-down searches) and iterate, keeping
the best candidate neighbors found at each step.  The binary tree used
in the treecode is built using the following rule: to split a node
$\alpha$, we compute its center ($\bx_c$), then the farthest point to
$\bx_c$ ($\bx_l$), then the farthest point to $\bx_l$ ($\bx_r$). We
project all the points on the line $(\bx_l,\bx_r)$, compute the
median, and split them into two groups. We recurse until every leaf
gets no more than $\ppl$ points. 

{\bf Node neighbor lists.} During the skeletonization, we need to
sample the far field. To do this we need to construct {\em node
  neighbor} lists. These lists are defined in \algref{a:neighbors}
and are constructed in a bottom-up fashion using a standard preorder
traversal of the tree. 
\begin{algorithm}[ht] 
\caption{\cfun{BuildNeighbors}($\alpha$)}\label{a:neighbors}
\begin{algorithmic}[1]
\STATE {\tt if} \cfun{IsLeaf}$(\alpha)$,  $\nla:= \left(\suu_{i\in\XXa}\nli\right) \sdf \XXa$ 
\STATE {\tt
 else} $\nla:= \left(\nl_{\rca} \suu \nl_{\lca}\right)\sdf\left(\XX_{\rca}\suu\XX_{\lca}\right)$
\end{algorithmic}
\end{algorithm}
The set-difference operations can be done in $\bigO( \log(N/m) )$ time\footnote{If $\log(N/m)$ bits is less than the size of an instruction, this can 
be done in constant time.}
per point using the binary-tree Morton ID of every node and every
point. The Morton ID is a bit array that codes the path from the root
to the node or point. The Morton ID of a point is the Morton ID of the
leaf node which contains it.

\textbf{Skeletonization of leaves.} Let $\XXa$ be the set of
points. Let the contribution of this node to all $\XX\sdf\XXa$ be
denoted as $\bKa$.  As noted above, we will approximate $\bKa$ by
computing a low-rank approximation using an inexact ID that is based
on sampling $\bKa$ to create a matrix $\bK(\XTa,\XXa)$ for some small
set of rows $\XTa$. Sampling the right points makes a difference and
can be expensive. In \cite{march-biros14}, we developed a sampling
scheme that is a hybrid between uniform sampling combined with nearest
neighbor sampling (for distance decaying kernels).  That is we choose
the nearest neighbors of the points in $\XXa$ which are not themselves
in $\XXa$ and then add uniformly sampled (without replacement) distant
points as needed to capture the far field.
\begin{equation}\label{e:sk_targets}
  \XTa = \nla\ \suu\ \mcfun{Sample}(\ \XX \sdf (\XXa \suu \nla),\
  \ssize-\card{\nla}\ ),
\end{equation}
That is, we randomly sample $\ssize-\card{\nla}$ points excluding the
points in $\alpha$ and we use these points along with the neighbors
$\nla$ as target points. We then compute the ID of $\bK(\XTa,\XXa)$ to
obtain the skeleton $\Sa$ and skeleton weights $\tilde{w}_\alpha$ for
$\bKa$.

On the other hand, if $\card{\nla} > \ssize$, we truncate $\nla$ to only include
$\ssize$ neighbors. We sort the points in $\nla$ by the distance from their 
nearest neighbor in $\alpha$, and keep the $\ssize$ closest points. 

\begin{algorithm}[ht] 
\caption{\cfun{Skeletonize}($\alpha$)}\label{a:skeletonize}
\begin{algorithmic}[1]
\STATE {\tt if} $\neg$ \cfun{IsLeaf}$(\alpha)$ 
\STATE \gap \cfun{Skeletonize}$(\rca)$,  \cfun{Skeletonize}$(\lca)$
\STATE \gap $\XXa= \MA{S}_\rca \suu \MA{S}_\lca$
\STATE Create sampling targets using~\eqref{e:sk_targets}
\STATE Skeletonize $\XXa$ using QR factorization and store $\Sa$ and $\tilde{w}_\alpha$
\end{algorithmic}
\end{algorithm}

\textbf{Skeletonization of internal nodes.} To build the far
field approximation for an interior node $\alpha$ we use the same algorithm.
Instead of using all the points  in the leaf
descendants of $\alpha$, we use the combined skeleton points $\SK_{\rca}
\suu \SK_{\lca}$ and the neighbors list $\nla$ constructed with
\cfun{BuildNeighbors}$(\alpha)$. Let $\tilde{w}_{\lca}$ and $\tilde{w}_{\rca}$ 
be the vectors of skeleton weights of the children.  

We compute the ID of the matrix $\bK(\XTa,\SK_{\rca} \suu \SK_{\lca})$
to obtain a skeleton for $\alpha$ and $P_\alpha$.  We then apply
$P_\alpha$ to the unskeletonized part of the concatenation of
$\tilde{w}_{\lca}$ and $\tilde{w}_{\rca}$ to obtain the skeleton
weights.  These ideas are summarized in
\cfun{SkeletonizeNode}$(\alpha)$.

%

{\bf Pruning.} During evaluation phase we use a standard top down
traversal. For every $\bx_i$, we start at the root and traverse the
tree. The pruning is based on the {\em neighbors} of $\bx_i$ and has
nothing to do with distance or kernel evaluations. Node $\alpha$ is
{\bf not pruned} if it is either an ancestor of $\bx_i$ or it is an ancestor of
{\em any} of the nearest neighbors of $\bx_i$:
\begin{equation}\label{e:prune}
\mcfun{Prune}(\alpha,i) = \mcfun{IsTrue}(\nexists j \in \{i \suu \nli\} : \alpha \in \anc_j)
\end{equation}
Note that the ancestor check can be done efficiently using
Morton IDs.  If \mcfun{Prune}$(\alpha,i)$ is true, we evaluate the
kernel at $\bx_i$ using the skeleton points and the equivalent weights
and do not traverse the children of $\alpha$.

\begin{algorithm}
\caption{$u_i =$ \cfun{Evaluate}$(\bx_i,\alpha)$}\label{a:down}
\begin{algorithmic}[1]
  \STATE {\tt if} \cfun{prune}$(\alpha,i)$, {\tt return} $K(\bx_i, \Sa){\b \ws}_\alpha$
  \hfill \COMMENT{Approximate (\eqref{e:prune})}
  \STATE {\tt if} \cfun{IsLeaf}$(\alpha)$, {\tt return}
  $K(\bx_i,\XXa){\b w}_\alpha$ \hfill \COMMENT{Direct evaluation}
  \STATE {\tt return} $\mcfun{Evaluate}(\bx_i,\rca) +  \mcfun{Evaluate}(\bx_i,\lca)$ \hfill\COMMENT{Recursion}
\end{algorithmic}
\end{algorithm}
\begin{algorithm}
\caption{$\b u$=\cfun{\ASKIT}$(\XX,\b w,\ns,\ssize,\ppl,\nl(\XX))$} \label{a:askit}
\begin{algorithmic}[1]
\STATE $\alpha =$\cfun{BinaryTree}$(\XX,\ppl)$ \hfill \COMMENT{Build  binary tree, $\alpha$ is the root}
\STATE \cfun{BuildNeighbors}$(\alpha)$ \hfill\COMMENT{Bottom-up traversal}
\STATE \cfun{SkeletonizeNode}$(\alpha)$ \hfill\COMMENT{Bottom-up traversal}
\STATE $u_i=\mcfun{Evaluate}(\bx_i,\alpha)\quad \forall i \in \XX$  \hfill\COMMENT{Top-down traversal}
\end{algorithmic}
\end{algorithm}
The evaluation algorithm and the overall scheme are summarized
in~\algref{a:down} and \algref{a:askit} respectively. 

To illustrate the difference of the proposed pruning compared to
standard distance pruning we conducted a numerical experiment in which
we compare the two approaches. The distance pruning criterion is
implemented as follows. Given a node $\alpha$ with points $\XXa$ we
compute its centroid $c_\alpha$ and a radius $R_\alpha =
\max_{x\in\XXa}\|x-c_\alpha\|_2$. Then given a target point $x$, we
prune if $\|x-c_\alpha\|_2>R$. Notice that in practice $R$ has to be
scaled to create some separation between the target and source points,
which makes pruning even harder. In Table~\ref{t:prune}, we report
the average number of nodes visited during the tree traversal for
evaluating the potential at several target points for a dataset of
$N=65,536$ points for different point distributions in 2D, 4D, 32D,
and 256D. We used a Gaussian distribution (intrinsic dimension is the
same as the ambient dimension), points distributed on a curve
(intrinsic dimension is one)\footnote{The equation of the curve is
  $x_i=f(i \pi t),\ t\in[0,1],\ i=1,\ldots,d$, with $f=\cos$ for odd
  $i$ and $f=\sin$ for even i.} and points uniformly distributed on a
hypersphere (intrinsic dimension is four).  These empirical results
show that distance based pruning is not possible in high dimensions
even if the underlying intrinsic dimension is small.

\begin{table}
\begin{center}
\begin{tabular}{@{}l| g r  g r  g r  g r}
\hline
{\bf dataset}      & \multicolumn{2}{|c|}{2D} & \multicolumn{2}{|c|}{4D} & \multicolumn{2}{|c|}{32D} & \multicolumn{2}{|c}{256D} \\
\hline
     &  $R$ &  $\kappa$ & $R$ &  $\kappa$  &  $R$ & $\kappa$  &  $R$ & $\kappa$ \\
\hline
 {\em curve}        &     1\% & 1\%          & 4\% & 1\%  &  30\% & 5\%    &  70\%&10\% \\
  {\em Gaussian}    &     4\% & 2\%          &  8\% & 4\%   &  100\% & 27\%  &  100\%& 30\% \\
  {\em hypersphere} &      -    &  -         & 26\% & 5\%  &  30\% & 6\%    &   32\%& 7\% \\
\hline
\end{tabular}
\end{center}
\caption{Here we demonstrate the difficulty of distance based pruning
  in high dimensions. Here {\bf dataset} indicates the dataset type,
  which we test in 2, 4, 32, and 256 dimensions.  The {\em spiral
    curve} has intrinsic dimensionality one, the {\em hypersphere} has
  intrinsic dimensionality four, and the {\em Gaussian} has the same
  intrinsic dimensionality as the ambient space. The $R$ column
  indicates distance based pruning and the $\kappa$ column indicates
  near-neighbor based pruning. The results indicate average values of
  nodes visited (as a percentage of the total number of nodes in the
  tree) during the downward pass averaged across target points. We
  used $\ppl = 64$ and $\kappa=128$. The larger the number
  of nodes visited, the less pruning takes place. This results indicate
  that simple distance based pruning is not effective in high dimensions. }
\label{t:prune}
\end{table}

\msection{Complexity and error}\label{s:complexity-and-error} %
\ASKIT{} has the following  parameters that control its computational 
cost and its accuracy. 
\begin{itemize}
\item $\ppl$: the number of points per leaf node; it controls the
  error and the runtime since it governs the trade-off between near
  and far interactions.
\item $\kappa$: the number of nearest neighbors for sampling and
  pruning. The larger $\kappa$ the less we prune. The larger $\kappa$
  the better our sample is when we compute the interpolative
  decomposition. If $\kappa$ is too large the computation becomes
  quite expensive.   In our tests, we set $\kappa=2\ppl$ and we have
  found that increasing $\kappa$ further does not improve the
  accuracy. 
\item $\ns$: the skeleton size. In essence, this is the rank we use for
  the far-field approximation. The higher $\ns$, the more accurate and
  expensive the skeletonization and evaluation phases are.  Here we
  fixed $\ns$ to be the same in all nodes.  A more efficient
  implementation will be to estimate  $\|K_R-K_SP\|$ and choose $\ns$
  adaptively to be different for each node. This is something that we
  are currently investigating. 
\item $\ssize$: the row sampling size for $\bKa$. Larger values allow
  a more accurate ID but slower skeletonization. We require
  $\ssize>\ppl$ and $\ssize>\ns$ so that ID problem is
  overdetermined. In our experiments we take
  $\ssize=\ns + 20$. In our experiments taking larger values
  does not increase the accuracy (if we keep everything else fixed).
\end{itemize}
So given the choices we describe above, there are two main parameters,
the number of points per box $\ppl$ and the skeleton size $\ns$.

{\bf Computational complexity.}  We assume that the nearest-neighbor
list $\nli$ for each point is given. Note that exact nearest-neighbors can 
be computed in $\bigO(N)$ time for low-intrinsic dimensional sets 
\cite{ram2009linear}
and approximation schemes, such as the one we use, are even faster. Thus, we 
only consider the cost of \ASKIT. 

 The number of leaves is $M= N/\ppl$ and the total number of nodes is
 $2M-1 = \bigO(M)$. We first consider the upward pass cost or
 skeletonization cost $T_S$. Then we consider the downward evaluation
 cost $T_E$. 

In the upward pass, the first calculation is the construction of
$\nla$ the per node neighbor lists that are used for sampling (given
by Algorithm~\ref{a:neighbors}). 

For leaf nodes, the cost of building $\nla$ (the per-node neighbor
lists) involves first merging the per-point neighbor lists $N_i$ for
all $i \in \XXa$, sorting them and removing duplicates, and then using
the Morton IDs to remove points that belong to a node. The complexity
of this operation per node is $\bigO(\kappa \ppl \log(\kappa \ppl ))$
and thus the total cost is $\bigO(N \kappa \log(\kappa \ppl) )$.  Once we have
$\nla$, we keep only the $\ssize$ nearest points to the $\XXa$ and use
them to construct the node ID. For internal nodes, the operation is
simpler. We simply merge the lists of the children, remove duplicates and 
points belonging to the node, then sort and truncate to $\ssize$ points (if 
necessary). The total complexity is $\bigO((N/m) \ssize \log \ssize)$. 

 The cost of skeletonization involves QR factorizations of $\bigO(M)$
 matrices of size $\ssize \times \ppl$ (at the leaves) or of size
 $\ssize \times (2 \ns)$ (for internal nodes). 
 In general, the time required for the evaluation of the kernel function 
 depends linearly on $d$.
 The QR factorization
 requires $\bigO(\ssize \ppl^2)$ for each of the $M$ leaves and 
 $\bigO(\ssize \ns^2)$ time for each of the $\bigO(M)$ internal nodes.
  Thus the total time to construct these matrices and compute their 
 factorizations is $\bigO(d M \ssize (\ppl^2 + \ns^2))$.
Therefore the
 total time $T_S$ for the upward pass (or skeletonization) is given by
\begin{equation}\label{e:time_skeleton}
T_S = \bigO(N \kappa \log(\kappa \ppl)) + \bigO((N/m)\, \ssize
\log\ssize) + \bigO(d N \ssize (\ppl^2 + \ns^2) / m).  
\end{equation}

The cost of the downward pass depends on our ability to prune.  Given
a target point $x_i$, let $\xi$ be the number of nodes we visit to 
compute $u_i$.  We decompose $\xi$ into two parts: $\xi_n$ for
nodes evaluated directly and $\xi_f$ for nodes approximated via the
skeleton ($\xi=\xi_f + \xi_n$). Given these parameters, the cost of
the downward pass for point $x_i$ is at most $\xi_n \ppl$ for direct
evaluation plus at most $\xi_f \ns$ for approximations.  Taking the
worse case values of $\xi_n$ and $\xi_f$ for all evaluation points,
the overall downward pass cost is bounded by $N (\xi_n \ppl + \xi_f
\ns)$. We now bound $\xi_n$ and $\xi_f$.

  For $\xi_n$, the worst case is that
for every point we visit $\kappa$ different leaves, (since we use the
point's nearest neighbors for pruning). Therefore
$\xi_n\leq\kappa$. Notice that this bound is quite pessimistic since
as $m$ increases the will be significant overlap of direct nodes for
target points belonging to the same leaf node. 
Additionally, the number of 
leaves visited will be smaller in the presence of a low intrinsic dimensional
structure. So this estimate is valid only for $m\ll N$.

To bound $\xi_f$ we proceed as follows.  Given a target point
$\bx_i$, \ASKIT{} will visit all nodes $\anc(\nli)$ and will prune all the remaining
nodes.  In the worst case, all the
elements of $\anc(\nli)$ are unique. Since $\card{\nli}=\kappa$,
$\card{\anc(\nli)}\leq\kappa \log M$. But if these nodes are unique,
that means we can prune their siblings by using their
skeletonization. Thus, $\xi_f=\bigO(\kappa \log M)$.  That is, we
visit $\kappa$ nodes assuming each neighbor of the evaluation belongs
to a different leaf node. 

Since the cost of evaluating the far field of a node to a point is
$\bigO(d \ns)$ (and assuming $m\ll N$) the overall complexity for
$T_E$ is given by
\begin{equation}\label{e:time_eval}
T_E = \bigO\left( d N \ppl\, \kappa + d N \ns\, \kappa\log\frac{N}{\ppl}
\right).
\end{equation}
The total cost of \ASKIT{} it $T=T_E +T_S$. 

Using the fact that in our implementation $\ssize = \bigO(\ns)$ and
$\kappa = \bigO(\ppl)$, the overall cost of \ASKIT{}  (for $m\ll N$)
\begin{equation}\label{e:time}
  T = \bigO\left( d N \ppl \left(\ns \ppl + \frac{\ns^3}{m} + \ns\log\frac{N}{\ppl}\right) \right).
\end{equation}
The ambient dimension $d$ enters only in the cost of kernel
evaluations. 
 
{\bf Error analysis.} There are two sources of error related to
$\bKa$: the far-field low rank approximation and the error in
computing this low rank approximation. This error appears in computing
the interpolative decomposition factorization due to the selection of a
subsample of the rows (the sampling to construct the skeleton).  It
can be shown that a particular importance sampling distribution (based
on what is known as statistical leverage scores) can provide a
$\bigO(\sigma_{\ns+1})$ reconstruction error if the sample size
$\ssize$ is proportional to $\ns \log \ns$ \cite{mahoney-drineas09}.
However, computing the leverage scores is more expensive than the
computation of the kernel summation exactly and thus, cannot be
applied in our context.

It is much cheaper to sample using a uniform distribution, but the
error bound is not as sharp. In~\cite{march-biros14}, we show that the
combined ID approximation and uniform sampling error in the far-field
from a single source node to the rest of the points can be bounded as
follows.  Let $K$ be the $n_\alpha \times m_\alpha$ matrix of interactions
between all the points $\XXa$ in the source node $\alpha$ and the
remaining $n_\alpha=N-m_\alpha$ points.%
\footnote{In addition we should exclude all the points not in $\alpha$ for
  which we use direct interactions.}
We sample $\ell$ rows (target points) uniformly and independently and
construct a rank $s$ interpolative decomposition of the sampled matrix
$\subK$ to select the skeleton size. 
Then, with high probability, the total 
error incurred is bounded by 
\begin{equation}
\|K - \subK \| \leq \zeta(n_\alpha,m_\alpha,\ns,\ell) \sigma_{\ns+1}(K). 
\label{eqn_full_error_bound}
\end{equation}	
where the quantity $\zeta$ is
\begin{equation}
\zeta(n_\alpha,m_\alpha,\ns,\ell) = 1+\sqrt{6\frac{n_\alpha}{\ell}} + \sqrt{1 + m_\alpha
    \ns (m_\alpha - \ns)}.
\label{eqn_zeta_def}
\end{equation}	
This result is Theorem 3.7 in \cite{march-biros14} with $\epsilon = 1/2$.  

For a given evaluation point, we incur this error each time we prune.
We know that the number of prunes is bounded by $\kappa
\log\frac{N}{m}$.  Denoting by $\MA{V}_i$ the nodes whose
skeletonization was used to evaluate the potential at $\bx_i$, and
using the fact that $m_\alpha=2^{\mathrm{level}(\alpha)}m$, the
overall absolute error is bounded by
\begin{equation}\label{e:error}
 |u_i - u_{\mathrm{exact}}(\bx_i)| \leq \|w\|\ \kappa
\log\frac{N}{\ppl}\ \max_{\alpha\in\MA{V}_i}\  \zeta\left(n_\alpha,
m_\alpha, \ns, \ell\right)\sigma^{\alpha}_{\ns+1}.
\end{equation}
This is a preliminary result. The remainder of the discussion is
qualitative. 

We remark that the ambient dimension does not appear in the error
estimate; $\kappa$ and errors in finding the exact nearest neighbors
affect the maximum of $\sigma^{\alpha}_{\ns+1}$; and the constant
$\zeta$ in \ref{eqn_zeta_def} depends on $n_\alpha$, $\ns$, and
$\ell$ as well as $\ppl_\alpha$.  This result is derived assuming the
skeletonization was computed using uniform random sampling. It is
rather pessimistic compared to a result derived for sampling using
leverage scores.  As we mentioned in our implementation, we employ a
heuristic in which we combine nearest neighbors and uniformly chosen
samples.  We also note that $\sigma_{\ns+1}^\alpha$ is bounded by the
$(\ns+1)^{\textrm{st}}$ singular value of the entire kernel
matrix. Therefore, in the case that the entire matrix is low rank, our
error bound will be small.

The parameter $\zeta$ can grow significantly if we keep the
skeleton size fixed because $\sigma_{\ns+1}^\alpha$ can grow and because
$\ppl_\alpha$ grows. To fix this, we can either increase $\ns$
adaptively or we can restrict the $\mathrm{level}(\alpha)$ by
starting the evaluation phase at a level which is a fixed distance
from the leaves. Also, in lower dimensions, a hybrid pruning rule that
combines distances and nearest neighbors could be used to derive
sharper error.

How does \eqref{e:error} compare to classical results for treecodes?
Using kernel specific analysis and distance pruning one can derive
analytic bounds for $\sigma_{\ns+1}$ based on the decay of
coefficients in analytic expansions. In that context $\ns$ corresponds
to the number of terms in the expansion. There is no explicit
dependence of $\zeta$ on $n_\alpha, \ppl_\alpha$ since the far field
is not computed algebraically or by target sampling.  Given that
distance pruning is critical in deriving those bounds, it is difficult to relate them to \ASKIT{} since it does not use distance pruning.  A
critical distance is the distance between the target point $\bx_i$ and
its $(\kappa+1)_\mathrm{st}$ nearest neighbor. (Recall that the
interactions of $x_i$ with all its first $\kappa$ neighbors are
computed directly.) The $(\kappa+1)_\mathrm{st}$ nearest neighbor may
end up being in a node that is pruned. Its distance to $\bx_i$
resembles the one used in the error estimates using distance pruning
and can be used to derive more quantitative bounds. Furthermore it can
be used to derive an adaptive (per point) selection of $\kappa_i$ to
further reduce the overall error of \ASKIT{}.

In summary, equation \eqref{e:error} is suggestive but not
particularly useful for quantitatively estimating the error and
choosing $\ns$ and $\kappa$. Information regarding the kernel and the
point distribution is required to derive a more precise estimate.
Next, we present an empirical evaluation of our scheme for
the Gaussian kernel with constant and variable bandwidth.

\msection{Experiments}\label{s:results}     \begin{table}[htp!]
\begin{center}
\begin{tabular}{@{}l| l l l| l g l l g r g l l@{}}
\hline
&\multicolumn{3}{c|}{\em parameters}&\multicolumn{3}{c}{\em
    errors}&\multicolumn{4}{c}{\em timings (secs)}&\multicolumn{2}{c}{\em pruning}\\\cline{8-11}
\rowcolor{gray!00}
{\bf run}& $\boldsymbol{\ppl}$& $\boldsymbol{\ns}$&
$\boldsymbol{h}$&{\bf hr}& $\boldsymbol{\epsilon}$&
      $\boldsymbol{\epsilon}_\kappa$& $\boldsymbol{T}_\kappa$&
$\boldsymbol{T}$& $\boldsymbol{T}_S$& $\boldsymbol{T}_E$&{\bf near}& ${\bf far}$\\ 
\hline
&\multicolumn{12}{l}{\em\color{gray!75} 4D Normal distribution,
  $N=100,000$, \quad $T_{\mathrm{direct}}=16$ secs}\\ 
1 &  64&   4& 0.21& 97& 5E-02& 3E-01& 5     & 5     & 1     & 4     & 15\%& 3 \%\\ 
2 &  64&  32& 0.21& 97& 1E-02& 3E-01& 5    & 7     & <1     & 6     & 15\%& 3 \%\\ 
3 & 256& 128& 0.21& 87& 2E-03& 8E-02& 7     & 8     & 1     & 7     & 5 \%& 1 \%\\ 
\hdashline[0.5pt/2pt]
4 &  64&   4& 1.00& 97& 7E-01& 1E+00& 5     & 4     & <1     & 4     & 15\%& 3 \%\\ 
5&  64&  32& 1.00& 97& 6E-02& 1E+00& 5     & 7     & <1     & 6     & 15\%& 3 \%\\ 
6& 256& 128& 1.00& 88& 3E-03& 9E-01& 7    & 8     & 1     & 7     & 5 \%& 1 \%\\ 
\hdashline[0.5pt/2pt]
7&  64&   4& 5.00& 97& 8E-01& 1E+00& 5     & 5     & <1     & 4     & 15\%& 3 \%\\ 
8&  64&  32& 5.00& 97& 1E-04& 1E+00& 5     & 7     & <1     & 6     & 15\%& 3 \%\\ 
9& 256& 128& 5.00& 88& 4E-08& 1E+00& 7     & 8     & 1     & 7     & 5 \%& 1 \%\\ 
\hline
&\multicolumn{12}{l}{\em\color{gray!75} 4D Normal distribution, $N=1,000,000$, \quad $T_{\mathrm{direct}}=1591$ secs}\\ 
10&  64&   4& 0.16& 95& 1E-01& 5E-01& 56    & 37    & 1     & 36    & 14\%& 3 \%\\ 
11 &  64&  32& 0.16& 95& 3E-02& 5E-01& 54    & 71    & 2     & 69    & 14\%& 3 \%\\ 
12 & 256&  32& 0.16& 84& 7E-03& 3E-01& 60    & 81    & 3     & 78    & 5 \%& 1 \%\\ 
&\multicolumn{12}{l}{\em\color{gray!55} Inexact neighbor information}\\ 
13 &  64&   4& 0.16& 26& 3E-01& 8E-01& 31    & 63    & 2     & 61    & 25\%& 4 \%\\ 
14 &  64&  32& 0.16& 26& 5E-02& 8E-01& 31    & 119   & 3     & 117   & 25\%& 4 \%\\ 
15 &  64& 128& 0.16& 27& 3E-02& 8E-01& 31    & 262   & 4     & 257   & 25\%& 4 \%\\ 
\hline
&\multicolumn{12}{l}{\em\color{gray!75} 16D Normal distribution, $N=1,000,000$, \quad $T_{\mathrm{direct}}=1630$ secs}\\ 
16  &  64&  32& 0.45& 35& 1E-03& 3E-03& 57    & 573   & 6     & 568   & 83\%& 23\%\\ 
17 &  64& 128& 0.45& 34& 7E-04& 4E-03& 57    & 725   & 7     & 717   & 83\%& 23\%\\ 
18 & 256& 128& 0.45& 26& 1E-04& 2E-03& 63    & 997   & 10    & 987   & 55\%& 8 \%\\ 
\hdashline[0.5pt/2pt]
19  &  64&  32& 1.76& 35& 7E-02& 1E+00& 57    & 564   & 6     & 559   & 83\%& 23\%\\ 
20 &  64& 128& 1.76& 35& 5E-02& 1E+00& 57    & 729   & 7     & 723   & 83\%& 23\%\\ 
21 & 256& 128& 1.76& 25& 2E-02& 1E+00& 63    & 1011  & 11    & 1000  & 55\%& 8 \%\\ 
\hline
&\multicolumn{12}{l}{\em\color{gray!75} 64D Normal distribution, $N=1,000,000$, \quad $T_{\mathrm{direct}}=1829$ secs}\\ 
22 &  64& 128& 0.75&  8& 3E-15& 3E-15& 63    & 984   & 10    & 974   & 99\%& 40\%\\ 
23 &  64& 128& 2.62&  8& 2E-01& 1E+00& 63    & 985   & 11    & 975   & 99\%& 40\%\\ 
24 &  64& 128& 4.98&  8& 8E-03& 1E+00& 63    & 991   & 10    & 981   & 99\%& 40\%\\ 
\hline
&\multicolumn{12}{l}{\em\color{gray!75}4D Normal distribution, variable $h$, $N=1,000,000$, \quad $T_{\mathrm{direct}}=1622$ secs}\\ 
25 &  64&  32& 1.00& 95& 3E-02& 1E+00& 55    & 145   & 2     & 143   & 14\%& 3 \%\\ 
26 &  64&  32& 5.00& 95& 4E-06& 1E+00& 57    & 144   & 2     & 142   & 14\%& 3 \%\\ 
\end{tabular}
\end{center}
\caption{Performance of \ASKIT{}:Here "$\boldsymbol{\ppl}$" is points
  per leaf, "$\boldsymbol{\ns}$" is number of skeleton points,
  "$\boldsymbol{h}$" the kernel bandwidth, "{\bf hr}" is the
  estimated percentage of correct neighbors. The relative error
  "$\boldsymbol{\epsilon}$" is estimated by direct
  evaluation on 10K randomly selected points;
  "$\boldsymbol{\epsilon}_\kappa$" is the error if we only use the
  near field. We report several timings: "$\boldsymbol{T}_\kappa$" is
  the time to construct the nearest neighbor lists, "$\boldsymbol{T}$"
  is the overall time of \ASKIT," $\boldsymbol{T}_S$" is
  skeletonization time, "$\boldsymbol{T}_E$" is evaluation time. To
  illustrate~\eqref{e:time}, we report the number of tree nodes
  visited per point during evaluation. "{\bf near}" is the average
  number of leaves visited as a percentage of the worst case $\kappa$;
  "{\bf far}" the number of nodes whose skeleton was used to evaluate
  the far field as a percentage of the worst case $\kappa
  \log(N/\ppl)$.  Also "$\boldsymbol{T}_{\mathrm{dir}}$" is the
  estimated time for a direct $N^2$ evaluation. All times are in
  seconds. We highlight the error, the treecode time, and the
  evaluation time. "{\bf run}" indexes the experiments.  }
\label{t:results0}
\end{table}

We present results for the Gaussian kernel
\[
K(\bx_i,\bx_j)=\exp(-\|\bx_i-\bx_j\|^2_2/h_j^2)
\]
 with constant and variable $h$; we select $h$ based on kernel density
estimation theory (Eq. 4.14 in ~\cite{silverman86}), so that
$h \propto N^{-1/(d+4)}$. The bandwidth $h$ plays a critical role in
assessing the accuracy of a treecode. For small and large $h$ the
kernel compresses well. But for certain $h$, which depends on the
underlying point distribution, the Gaussian kernel does not compress
well.  The base value we use here is optimal for a constant-width
Gaussian kernel and normally distributed points. Also let us remark
that doing a simple sweep for $h$ may miss the values of $h$ for which
$K$ fails to compress. (More discussion and results can be found
in~\cite{march-biros14}.)  In our experiments, we choose $h$ large
enough so that far-field is necessary for accurate summation and small
enough so that the kernel is difficult to compress.

\ASKIT{} has been implemented in C++. The direct evaluation is highly
optimized using BLAS but the other parts of the code are proof of
principle implementations.  We use the Intel MKL for linear algebra,
and use OpenMP and MPI parallelism in all phases of the algorithm. 

The experiments were carried out on the
Stampede system at the Texas Advanced Computing Center. Stampede
entered production in January 2013 and is a high-performance Linux
cluster consisting of 6,400 compute nodes, each with dual, eight-core
processors for a total of 102,400 CPU-cores. The dual-CPUs in each
host are Intel Xeon E5 (Sandy Bridge) processors running at 2.7GHz
with 2GB/core of memory and a three-level cache. The nodes also
feature the new Intel Xeon Phi coprocessors. Stampede has a 56Gb/s
FDR Mellanox InfiniBand network connected in a fat tree configuration
which carries all high-speed traffic (including both MPI and parallel
file-system data).

To test \ASKIT{}, we use normally distributed points in 4, 16, and 64
dimensions for which the intrinsic and ambient dimension coincide. We
present results for 100K and 1M points.  We also consider the
embedding in 1000D of a set of points normally distributed in a 4D
hypersphere. We also used a UCI ML repository dataset
(SUSY~\cite{bache-lichman13}) with 5M points in 18 dimensions.  In all
experiments, the sources and targets coincide, and we report timings
for all pairwise interactions.  We present wall-clock times for
finding the neighbors, constructing the skeletonization, and the
evaluation.  Let us remark that for problems that require multiple
all-to-all kernel evaluations (e.g., regression), the cost of
construction, skeletonization, and neighbor finding is amortized over
the iterations. The only per-iteration cost is $T_E$.

In~\tabref{t:results0} and \tabref{t:results1}, we report results that
show the feasibility of \ASKIT{}.  The performance of our nearest
neighbor search affects the overall runtimes and the performance
of \ASKIT. Although it is an independent component, we report the
numbers since nearest neighbors must be computed somehow.  For this
reason, we report the nearest neighbor hit rate accuracy (percentage
of correct neighbors) and the timings.  We test the accuracy of our
nearest-neighbors and \ASKIT{}, using exhaustive searches and direct
evaluations on 10K randomly sampled points.

\begin{table}[htp!]
\begin{center}
\begin{tabular}{@{}l| l l l| l g l l g r g l l@{}}
\hline
&\multicolumn{3}{c|}{\em parameters}&\multicolumn{3}{c}{\em
    errors}&\multicolumn{4}{c}{\em timings (secs)}&\multicolumn{2}{c}{\em pruning}\\\cline{8-11}
\rowcolor{gray!00}
{\bf run}& $\boldsymbol{\ppl}$& $\boldsymbol{\ns}$&
$\boldsymbol{h}$&{\bf hr}& $\boldsymbol{\epsilon}$&
      $\boldsymbol{\epsilon}_\kappa$& $\boldsymbol{T}_\kappa$&
$\boldsymbol{T}$& $\boldsymbol{T}_S$& $\boldsymbol{T}_E$&{\bf near}& ${\bf far}$\\ 
\hline
&\multicolumn{12}{l}{\em\color{gray!75} 1000D (4d-intrinsic), $N=1,000,000$, \quad $T_{\mathrm{direct}}=7315$ secs}\\ 
27 &  64&  32& 0.75& 99& 6E-02& 1E+00& 204   & 393   & 11    & 382   & 14\%& 3 \%\\ 
28 &  64&  32& 3.75& 99& 2E-06& 1E+00& 195   & 391   & 10    & 381   & 14\%& 3 \%\\ 
\hline
&\multicolumn{12}{l}{\em\color{gray!75} 18D (UCI SUSY), $N=5,000,000$, \quad $T_{\mathrm{direct}}=41,000$ secs}\\ 
29 &  64& 128& 0.40& 91& 1E-01& 7E-01& 709   & 1510  & 32    & 1478  & 43\%& 9 \%\\ 
30 &  64& 128& 1.00& 90& 7E-02& 1E+00& 567   & 1530  & 33    & 1497  & 43\%& 9 \%\\ 
31 &  64& 128& 5.00& 90& 1E-03& 1E+00& 577   & 1566  & 33    & 1533  & 43\%& 9 \%\\ 
\hline
&\multicolumn{12}{l}{\em\color{gray!75} MPI-parallel 128D (4d-intrinsic),
  $N=5,000,000$, \quad $T_{\mathrm{direct}}=280,000$ secs}\\ 
32 &  64& 128& 0.50& 100& 1E-03& 1E+00& 233   & 1339  & 12    & 1158  & 17\%& 9\%\\ 
33 &  64& 128& 0.50& 100& 1E-03& 1E+00& 43    & 170   & 1     & 144 & 17\%& 9\%\\ 
\hline
&\multicolumn{12}{l}{\em\color{gray!75} MPI-parallel 64D
  (4d-intrinsic) 512 nodes, (8,192 cores) $N=100,000,000$ }\\ 
34 &  64& 128& 0.37& 99& 1E-02& 1E+00& 137   & 1305  & 6     & 880   & 16\%& 19\%\\ 
\end{tabular}
\end{center}
\caption{Performance of \ASKIT{}: Here "$\boldsymbol{\ppl}$" is points
  per leaf, "$\boldsymbol{\ns}$" is number of skeleton points,
  "$\boldsymbol{h}$" the kernel bandwidth, "{\bf hr}" is the estimated
  percentage of correct neighbors. The relative error
  "$\boldsymbol{\epsilon}$" is estimated by direct evaluation on 10K
  randomly selected points; "$\boldsymbol{\epsilon}_\kappa$" is the
  error if we only use the near field. We report several timings:
  "$\boldsymbol{T}_\kappa$" is the time to construct the nearest
  neighbor lists, "$\boldsymbol{T}$" is the overall time of \ASKIT,"
  $\boldsymbol{T}_S$" is skeletonization time, "$\boldsymbol{T}_E$" is
  evaluation time. To illustrate~\eqref{e:time}, we report the number
  of tree nodes visited per point during evaluation. "{\bf near}" is
  the average number of leaves visited as a percentage of the worst
  case $\kappa$; "{\bf far}" the number of nodes whose skeleton was
  used to evaluate the far field as a percentage of the worst case
  $\kappa \log(N/\ppl)$.  Also "$\boldsymbol{T}_{\mathrm{dir}}$" is
  the estimated time for a direct $N^2$ evaluation. All times are in
  seconds. We highlight the error, the treecode time, and the
  evaluation time. "{\bf run}" indexes the experiments.  Runs (33-34)
  are done using 16 and 256 nodes respectively and the MPI
  library~\cite{mpi96}. }
\label{t:results1}
\end{table}

{\bf Discussion.} For low-accuracy approximations our scheme
outperforms the direct evaluation at about 1M points. Depending on the
kernel and the accuracy the speed-up can be less dramatic or the
cutoff may be at much higher $N$.
 Now we make some additional remarks
on our runs.  

In runs 1--9 we show how the method converges for different bandwidths
and values of $\ppl$ and $\ns$ for 100K points that are normally distributed
in 4D. Note that the error $\epsilon$ converges with increasing $\ppl$
and $\ns$. The convergence can be quite rapid (runs 7--9). In most of
the runs, the far-field is critical in getting accuracy. To show this,
we report $\epsilon_\kappa$, the error in $u$ constructed using only
the $\kappa=2\ppl$ nearest neighbors for each point.  We see that the
far field is essential, and truncation does not get a single digit
correct. On the other hand for run 22, with $h=0.75$ the far field is
wasted effort. In runs 23--24 the far field is essential.

 In runs 27-28, we consider a problem in 1000D. The
scheme converges quickly and it is 20$\times$ faster than the direct
evaluation. For the UCI dataset (29-31) \ASKIT{} is 25$\times$ faster
than the direct evaluation.  To demonstrate the effects of using the
nearest neighbors compare runs 10--12 to runs 13--15. The only
difference is that that we use a very approximate search so the
hit-rate "{\bf hr}" (correct neighbors/$\kappa$) small. As a result
the errors are higher and the pruning is not as effective (we visit
more leaves and more internal nodes). $T_E$ is almost 3$\times$
larger.  Finally, notice that we increase the dimension the neighbors
in general become less accurate. This is because we use a fixed number
of iterations in our greedy neighbor search. The skeletonization costs
are negligible compared to the evaluation costs. 

Runs (1-31) took place on a single node.  In runs (32-34) we show a
distributed memory run for 5M points in 128D on 16 and 256 nodes
resulting a 2000$\times$ speed-up over one-socket direct
evaluation. Our largest run on 512 nodes (8,192 cores) we evaluated
the sum for 100 million points in 64 dimensions. 


\msection{Conclusions}\label{s:conclusions} We presented a new scheme for high dimensional $N$-body problems
and conducted a proof-of-concept experimental study.  Our scheme is
based only on kernel evaluations, uses neighbor-based pruning, and
uses neighbor-sampled interpolative decomposition to approximate the
far field. Since this method is new, there are many open problems and several
opportunities for optimization. The most pressing one is deriving
 a rigorous error bound that incorporates our sampling scheme;
this is ongoing work. There is also further work to be done in
optimizing the performance of the scheme, in adaptive determination of
the skeleton size, and in improving the sampling. Finally, notice that if we
are given similarities and a hierarchical clustering, our scheme does
not involve any distance calculations so it should be possible to
apply to points (objects) in non-metric spaces.

\bibliographystyle{siam}
\bibliography{gb}
\end{document}